\documentclass[11pt]{article}
\pdfoutput=1
\usepackage[utf8]{inputenc}

\usepackage{jhep pub} 
\usepackage{hyperref}

\usepackage[T1]{fontenc} 

\usepackage{comment}

\usepackage{amsmath, amssymb}

\usepackage{mathtools}

\usepackage{graphicx}

\usepackage{bbold}

\usepackage{multirow}

\usepackage{slashed}

\usepackage{tikz}
\usetikzlibrary{decorations.pathmorphing}
\usetikzlibrary{decorations.markings}

\allowdisplaybreaks

\title{Two-loop four-point amplitudes on the Coulomb branch of 
${\mathcal{N}}=4$ super Yang-Mills
}
\author[a,b]{Wojciech Flieger,}
\author[c]{Johannes Henn,}
\author[c]{Anders Schreiber,}
\author[d]{Jaroslav Trnka.}

\affiliation[a]{Dipartimento di Fisica e Astronomia, Universit\`a degli Studi di Padova, Via Marzolo 8, I-35131 Padova, Italy}
\affiliation[b]{INFN, Sezione di Padova, Via Marzolo 8, I-35131 Padova, Italy}

\affiliation[c]{Max-Planck-Institut f\"{u}r Physik, Werner-Heisenberg-Institut, D-80805 M\"{u}nchen, Germany}

\affiliation[d]{Center for Quantum Mathematics and Physics (QMAP), University of California, Davis, 95616, California, USA}

\emailAdd{flieger@pd.infn.it, henn@mpp.mpg.de, ohrbergs@mpp.mpg.de, trnka@ucdavis.edu}

\preprint{MPP-2025-4}

\abstract{
We explore scattering amplitudes on the Coulomb branch of maximally supersymmetric Yang-Mills theory. 
We introduce a particular pattern of scalar vacuum expectation values that allow us to define amplitudes with a different mass pattern compared to what was studied previously.
This is motivated by an extension of the Amplituhedron that leads to infrared-finite four-particle amplitudes involving massive particles.
We work out the Feynman rules on the Coulomb branch and use them, together with generalized unitarity techniques, 
to perform consistency checks on the Amplituhedron expectations for the one- and two-loop integrands for the four-particle amplitude. 
We present details of the computation of the required two-loop four-point integrals via a four-dimensional version of the differential equations method. 
Finally, we study the Regge limit of the four-point amplitude, including the first power suppressed terms. 
We find that when organized in terms of a suitable expansion parameter, the subleading power term exponentiates, 
with the exponent matching the anomalous dimension of a cusped Wilson line with a local operator insertion. 
The latter is known from integrability, which leads to a prediction at higher loop orders in the Regge limit.
}

\begin{document}
\tikzset{->-/.style={thick,decoration={
  markings,
  mark=at position .5 with {\arrow{>}}},postaction={decorate}}}

\maketitle

\section{Introduction}

The maximally supersymmetric Yang-Mills theory ($\mathcal{N}{=}4$ sYM) in the planar limit (large number of colors, $N \rightarrow\infty$) has been an ideal testing ground for novel theoretical ideas. The study of scattering amplitudes has revealed many surprising properties, hidden symmetries, and new conceptual approaches, where Lagrangians and Feynman diagrams do not play a central role \cite{Arkani-Hamed:2022rwr}. This includes the exponentiation of infrared divergences and finite parts  \cite{Bern:2005iz}, the discovery of dual conformal symmetry \cite{Drummond:2007au,Drummond:2008vq}, symbols and bootstrap methods \cite{Goncharov:2010jf,Dixon:2011pw,Dixon:2011nj,Caron-Huot:2019vjl}, antipodal duality \cite{Dixon:2021tdw}, S-matrix flux tube approach \cite{Basso:2013vsa,Basso:2014hfa}, Grassmannian formulation and on-shell diagrams \cite{Arkani-Hamed:2010zjl,Arkani-Hamed:2009ljj,Arkani-Hamed:2012zlh,Arkani-Hamed:2014bca,Paranjape:2022ymg,Brown:2022wqr} and the Amplituhedron \cite{Arkani-Hamed:2013jha,Arkani-Hamed:2013kca,Arkani-Hamed:2017vfh,Damgaard:2019ztj,Ferro:2022abq,Herrmann:2022nkh}. The Amplituhedron provides a radically new picture for the particle interactions where the scattering amplitude is reproduced by the volume form associated with the geometric space. This has led to many new insights and results for amplitudes \cite{Franco:2014csa,Dennen:2016mdk,Arkani-Hamed:2018rsk,Herrmann:2020qlt,Dian:2022tpf,He:2023rou,Arkani-Hamed:2023epq,Ferro:2024vwn,De:2024bpk,Dian:2024hil}, including a novel amplitude decomposition using negative geometries \cite{Arkani-Hamed:2021iya,He:2022cup,Brown:2023mqi,Henn:2023pkc,Lagares:2024epo,Chicherin:2024hes,Glew:2024zoh}. The Amplituhedron has also been extensively studied from a mathematical perspective and is connected to recent developments in many disciplines, including cluster algebras or tropical geometry \cite{Galashin:2018fri,Lukowski:2020dpn,Parisi:2021oql,Even-Zohar:2023del,Akhmedova:2023wcf,Even-Zohar:2024nvw,Lam:2024gyg,Parisi:2024psm,Galashin:2024ttp}.

Loop amplitudes in the ${\cal N}=4$ sYM theory suffer from infrared divergences. The standard method is the use of the dimensional regularization where the calculation is made in $4-2\epsilon$ dimensions and the amplitude becomes an expansion in $1/\epsilon$. 
However, this procedure breaks many of the symmetries of ${\cal N}=4$ sYM, especially the hidden dual conformal symmetry.
The AdS/CFT correspondence suggests an alternative, symmetry-friendly way of obtaining finite amplitudes, by switching on the radial direction in AdS${}_{5}$ space, as done in the string theory-based computation of gluon scattering in strong coupling \cite{Alday:2007hr} and further used in \cite{Kawai:2007eg,McGreevy:2008zy,Berkovits:2008ic}. From a field theory perspective the mass regulator is introduced by moving away from the origin in the moduli space of vacua parametrized by the vacuum expectation values (VEV) of scalars of the theory, the so-called Coulomb branch. The masses in this picture are produced via the Higgs mechanism by spontaneous breaking of the gauge symmetry. 

A systematic study of Coulomb branch loop amplitudes in field theory was initiated in reference  \cite{Alday:2009zm}. There, it was argued that e.g. by choosing a vacuum expectation value that breaks the symmetry according to $U(N+M) \to U(N) \times U(1)^{M}$, one may define infrared-finite scattering amplitudes (in the large $N$ limit). Moreover, it was shown that the dual conformal symmetry holds, provided that the masses transform in a particular way.
The amplitudes enjoy a number of exact limits, for example in the high-energy limit and in Regge limits \cite{Alday:2009zm,Henn:2010bk}.
Higher-point generalizations were studied in ref. \cite{Henn:2010ir}. 
Recent studies of Coulomb branch physics include \cite{Loebbert:2020hxk,MdAbhishek:2023nvg,MdAbhishek:2023nvg,Belitsky:2024agy,Bork:2022vat,Ivanovskiy:2024vel}.
There are also intriguing observations about conjectural enhanced symmetries for loop integrands, which imply relations to correlation functions \cite{Caron-Huot:2021usw}.

It is interesting to study how massless sYM theory is related to its massive counterpart on the Coulomb branch. In reference \cite{Craig:2011ws}, the authors showed a tree-level connection between a soft scalar limit of the massless amplitudes and a small mass limit of the Coulomb branch amplitudes. Another interesting question concerns various kinematical limits of the Coulomb branch amplitudes. Particularly interesting is the Regge limit, where one of the kinematic invariants is sent to infinity. For the three-loop order it has been shown that in this limit the four-point amplitude exponentiates up to the subleading term \cite{Bruser:2018jnc}. The exponents are given by the leading and subleading Regge trajectories which can be related to the angle-dependent cusp anomalous dimension 
\cite{Henn:2010bk, Correa:2012nk}. An interesting interpretation  of the Coulomb branch setup in terms of a relativistic hydrogen-like system was provided in reference \cite{Caron-Huot:2014gia}.

In this paper, we study the connection between the Coulomb branch amplitudes and the Amplituhedron geometry. We initiated this effort in \cite{Arkani-Hamed:2023epq}, where we defined a deformed Amplituhedron that produced finite amplitudes. The infrared finiteness was achieved by introducing two deformation parameters. These parameters can be related to kinematic invariants and masses, providing a particular mass configuration of the four-point amplitude. In that reference it was also shown that this configuration can be obtained via a choice of  Coulomb branch VEV, different from that used in reference \cite{Alday:2009zm}. However, the connection between the deformed Amplituhedron and this field theory setup remains conjectural.

In the present paper, we provide details of the field theory setup and computation, up to two loops.
To determine the loop integrand and to evaluate the novel two-loop Feynman integrals, we closely follow references \cite{Alday:2009zm} and \cite{Caron-Huot:2014lda}, respectively, adapting the methods employed there to the present setup. 
In particular, a major advantage of dealing with finite loop integrals is that we can use the dedicated four-dimensional version of integration-by-parts (IBP) \cite{Chetyrkin:1981qh} and differential equations (DE) \cite{Kotikov:1990kg, Bern:1995db,Gehrmann:1999as,Henn:2013pwa} methods for Feynman integrals from reference \cite{Caron-Huot:2014lda}. This leads to significant simplifications, compared to standard approaches.
An additional advantage of this method is that it makes dual conformal symmetry manifest, thereby reducing the number of integrals that need to be considered. 

We provide a pedagogical, self-contained exposition of this method and apply it to the relevant one- and two-loop integrals. 
In this way, we obtain an analytic solution for the Coulomb branch amplitude up to the two-loop level. Interestingly, the result can be fully expressed in terms of classical polylogarithms.
With the analytic expression at hand, one can study interesting kinematic limits.
The Regge limit is of particular interest.

Previously, the dual conformal symmetry on the Coulomb branch was used to show that in planar sYM theory, the angle-dependent cusp anomalous dimension is the same as the (massive) Regge trajectory in that theory \cite{Henn:2010bk}. This generalizes a corresponding QCD relation, which, however, holds only in the massless limit.
Moreover, employing a dual conformal partial wave expansion, reference \cite{Bruser:2018jnc} found an intriguing exponentiation pattern also at the first power suppressed level in the Regge limit, with the exponent given by a Wilson loop with insertion of the scalar operator. This motivates us to investigate the Regge limit for the new VEV pattern considered here.

We use the analytic two-loop results to evaluate the Regge limit, including power suppressed terms. After taking into account the kinematic differences, we find an exponentiation pattern similar to that of reference \cite{Bruser:2018jnc}.
Very interestingly, the subleading power term again exponentiates exactly (at least, up to the order studied), but with a different exponent: the anomalous dimension of a cusped supersymmetric Wilson loop, with a scalar insertion (which is the same scalar to which the Wilson loop couples). This is particularly exciting, as this anomalous dimension is known from the integrability \cite{Correa:2012hh,Drukker:2012de,Gromov:2013qga}. Generalizing the VEV breaking pattern to include an additional, SO(6), angle between the scalar fields of sYM, we are able to match the Regge limit exponents to locally supersymmetric cusped Wilson lines, where the SO(6) angle corresponds to the difference in scalar coupling between the two Wilson line segments \cite{Drukker:2011za}.

The paper is organized as follows. In Section \ref{sec:setup}, we first review the deformed Amplituhedron four-point amplitudes of reference \cite{Arkani-Hamed:2023epq}. We then discuss and compare several configurations of vacuum expectation values in $\mathcal{N}=4$ sYM, one of which matches the Amplituhedron integrands, at least to the two-loop order.
In Section \ref{sec:de}, we employ four-dimensional differential equations in order to compute the massive one- and two-loop four-point integrals. We then use those results, in Section \ref{sec:Regge}, to investigate the Regge limit of the amplitudes. We find a pattern of exponentiation, including the subleading power terms. This is similar to observations in reference \cite{Bruser:2018jnc}, but involves a different anomalous dimension compared to that reference. The paper is accompanied by Appendix \ref{sec:appendix} which contains formulas for changing between different parametrization of the kinematics.


\section{The Amplituhedron and Coulomb branch amplitudes}
\label{sec:setup}

\subsection{Four-point amplitudes from a deformed Amplituhedron}
\label{subsec:Amplituhedron}

The object in which we are interested is the four-point amplitude associated with the deformed amplituhedron \cite{Arkani-Hamed:2023epq}. 
Its perturbative expansion takes the form
\begin{equation} \label{eq:amplitudeDeformed}
M_{A} = 1 + g^{2} M^{(1)}_{A} + g^4 M^{(2)}_{A} + {\cal{O}}(g^6) \, ,   
\end{equation}
where 
\begin{align}\label{eq:defg}
    g^2 = {g_{\rm YM}^2 N}/({16 \pi^2}) \,,
\end{align} with $g_{\rm YM}$ being the Yang-Mills coupling. It was shown in ref. \cite{Arkani-Hamed:2023epq} that the Amplituhedron conditions fix the integrands for 
$M^{(1)}_{A}$ and $M^{(2)}_{A}$ up to some normalization constants, which we denote by $n^{(1)}, n^{(2a)}, n^{(2b)}$ below.

In that reference, the integrands were written in momentum twistor space.
The amplitude $M_{A}$ is dual conformal invariant, which implies that it depends on two dual conformal cross ratios $u$ and $v$, i.e. $M_{A}=M_{A}(u,v)$, which are naturally expressed in terms of momentum twistors \cite{Arkani-Hamed:2023epq}.
For convenience of readers not familiar with twistor variables, we present the one- and two-loop amplitudes in momentum space. At one loop, the answer can be written in terms of a certain box integral,
\begin{align}
M^{(1)}_{A}(u,v) =& n^{(1)}  I^{(1)}(u,v) \,,
\end{align}
with\footnote{Note that we use the metric $(+---)$, as opposed to the choice made in reference \cite{Alday:2009zm}.}
\begin{align}\label{eq:oneloopboxmomentum}
\begin{split}
I^{(1)}(u,v) =& \, \int \frac{d^{4}k}{i \pi^2}  \frac{{\left[-s+(m_{1}-m_{3})^{2}\right]\left[-t + (m_{2}-m_{4})^{2}  \right]}}{[ -k^2+m_1^2][-(k+p_1)^2+m_2^2]}  \\
& \qquad\quad \times\frac{1}{ [-(k+p_{1}+p_2)^2+m_3^2][-(k+p_{1}+ p_2+p_3)^2+m_4^2]   } \,.
\end{split}
\end{align}
Here, the kinematics is as follows,
\begin{align}
p_1^2 = m_1^2 +m_2^2 \,, \quad p_2^2 = m_2^2+m_3^2
\,, \quad p_3^2 = m_3^2+m_4^2
\,, \quad p_4^2 = m_4^2+m_1^2
\,.
\end{align}
As alluded to above, the integral in eq. (\ref{eq:oneloopboxmomentum}) is invariant under dual conformal symmetry \cite{Alday:2009zm}.
The two dual conformal invariants from \cite{Arkani-Hamed:2023epq} take the form
\begin{align}\label{eq:defdualconformalinvariantsuv}
u = \frac{(m_1^2+m_3^2-s)^2}{4 m_1^2 m_3^2} \,,\qquad v= \frac{(m_2^2+m_4^2-t)^2}{4 m_2^2 m_4^2} \,.
\end{align}
Because of the dual conformal symmetry we can choose special configurations without a loss of generality. For example, setting $m_i = m$, we still have a genuine function $M^{(1)}_{A}$ of two variables $u=[1-s/(2 m^2)]^2$, $v=[1-t/(2 m^2)]^2$; this function is then represented by a box integral with uniform internal mass $m$, and on-shell conditions $p_i^2 = 2 m^2$.

At two loops, the answer is given by the sum of two double box integrals,
\begin{align}\label{eq:expectedtwoloopanswer}
M^{(2)}_{A}(u,v) = n^{(2a)} I^{(2)}(u,v) + n^{(2b)} I^{(2)}(v,u) \,,
\end{align}
where
\begin{align}
\begin{split}
    I^{(2)}(u,v) &=  \int \frac{d^{4}k_{1}}{i \pi^{2}} \frac{d^{4}k_{2}}{i \pi^{2}} \frac{{\left[-s+(m_{1}-m_{3})^{2}\right]^{2} \left[-t + (m_{2}-m_{4})^{2}  \right]}}{[-k_{1}+m_{1}^{2}][-(k_{1}+p_{1})+m_{2}^{2}][-(k_{1}+p_{1}+p_{2})^{2}+m_{3}^{2}]} \\
&\hspace{-1.5cm} \times \frac{1}{[-(k_{1}-k_{2})^{2}][-k_{2}+m_{1}^{2}][-(k_{2}+p_{1}+p_{2})+m_{3}^{2}][-(k_{2}+p_{1}+p_{2}+p_{3})+m_{4}^{2}]} \,.
\end{split}
\end{align}
One of the goals of the present paper is to test the conjecture made in \cite{Arkani-Hamed:2023epq} that the deformed Amplituhedron amplitude of eq. (\ref{eq:amplitudeDeformed}) can be obtained from a Coulomb branch setup,
and to determine the coefficients $n^{(1)}, n^{(2a)}$, and $n^{(2b)}$.

\subsection{Coulomb branch setup}
\label{subsec:setup_coulomb}

We consider the breaking of $U(N + M)$ symmetry to $U(N) \times U(1)^{M}$, following very closely the derivation in reference \cite{Alday:2009zm}. 
The action of the Coulomb branch theory is given by \cite{Brink:1976bc} 
\begin{align}
\hat{S}_{\mathcal{N}=4}^{U(N+M)} = \int d^4 x \text{Tr} \bigg( - \frac{1}{4} \hat{F}^2_{\mu \nu} - \frac{1}{2} (D_\mu \hat{\Phi}_I)^2 + \frac{\tilde{g}^2}{4} [ \hat{\Phi}_I, \hat{\Phi}_J]^2 + \frac{i}{2} \hat{\bar{\Psi}} \Gamma^\mu D_\mu \hat{\Psi} + \frac{\tilde{g}}{2} \hat{\bar{\Psi}} \Gamma^I [\hat{\Phi}_I , \hat{\Psi}]   \bigg) \, ,
\end{align}
{where $\tilde{g}= g_{YM}/\sqrt{2}$}, 
$D_\mu = \partial_\mu - i \tilde{g} [\hat{A}_\mu, \cdot ]$ and $\hat{A}_{\mu}$ are vector fields, $\hat{\Psi}$ are Majorana-Weyl 
spinors and $\hat{\Phi}_{I}$ are the six scalars of the theory (with $I=4,\hdots, 9$).
All fields of the theory can be represented as Hermitian matrices.
Introducing the collective notation $\mathcal{\chi}$ for a generic field, ${\cal \chi} \in \{ A , \Phi, \Psi\}$, we can describe the $(N+M) \times (N+M)$ block structure of these matrices as follows, 
\begin{align}
    {\chi} = \begin{pmatrix}
    \chi_{a,b} & \chi_{a \, N+j} \\
    \chi_{N+i, a} & \chi_{N+i, N+j} 
    \end{pmatrix} \, .
\end{align}
where $a, b = 1 , \ldots, N$, and $i,j = 1 , \ldots, M$. 
We will pay particular attention to the scalars, as we give a VEV to them in the $M\times M$ part of the matrix
\begin{align}
  (\Phi_I)_{N+i, N+j}  \longrightarrow (\Phi_I)_{N+i, N+j} +   \frac{1}{\tilde{g}} (M_{I})_{ij}  \,.
\end{align}
and 
\begin{align}
M_{{I}} = \textrm{diag}( \vec{n}_{1} m_{1}\,, \ldots \,,\vec{n}_{M}m_{M}) \,,   
\label{eq:vev_matrix}
\end{align}
where $\vec{n}_{i}$ are $SO(6)$ vectors. 
Depending on which part of the $(N+M) \times (N+M)$ block we are considering, the fields acquire different masses, as we discuss in the following specific cases.

\subsubsection*{VEV breaking pattern considered in \cite{Alday:2009zm} (AHPS).}

In \cite{Alday:2009zm} the VEV matrix was chosen to be non-zero along the same $SO(6)$ directions for all $i$,
\begin{align}
\vec{n}_{i} = \delta_{I9} \,, \quad i= 1, \hdots , M \,, 
\end{align}
with general values of the mass parameters $m_i$.
This results in three types of fields:
\begin{itemize}
\item Massless fields $\mathcal{\chi}_{ab}$
 from the unbroken $N \times N$ part of the gauge group
\item `Light' fields  $\mathcal{\chi}_{N+i, N+j}$
with masses $|m_{i} - m_{j}|$, from the off-diagonal $N \times M$ and $M\times N$ parts of the matrices
\item  `Heavy' fields $\mathcal{\chi}_{N+i, a}$ with masses $m_{i}$, from the $M \times M$ part of the matrices
\end{itemize}
In the large $N$ limit, scattering `light' fields is mediated by heavy fields, resulting in infrared-finite scattering amplitudes \cite{Alday:2009zm}.
A particularly interesting case is $m_{i}=m$.  
Let us also mention that dual conformal symmetry, which corresponds to isometries of AdS${}_{5}$ in the dual string theory, allows one to relate different mass configurations \cite{Alday:2009zm}.

\subsubsection*{VEV breaking pattern considered in \cite{Arkani-Hamed:2023epq} (AFHSTa)}

In the present paper, for simplicity of presentation, we restrict ourselves to $M=4$.
We consider the following configuration \cite{Arkani-Hamed:2023epq} 
\begin{align}
M_{I} =\textrm{diag}(m_{x}\delta_{I8},-m_{y}\delta_{I9},-m_{x}\delta_{I8},m_{y}\delta_{I9}) \,,     
\label{eq:vev_deformed}
\end{align}
which can be decomposed into two independent $SO(6)$ directions $I=8$ and $I=9$. 
Thus, the non-zero VEVs are represented by the following $(N+M)\times(N+M)$ matrices,
\begin{align}
\begin{split}
\langle \Phi_8 \rangle
&=
 \frac{m_{x}}{\tilde{g}}\text{diag}(\overbrace{0, \ldots 0}^{N},1,0,-1,0)  \,,\\
\langle \Phi_9 \rangle
&= 
\frac{m_{y}}{\tilde{g}}\text{diag}(\overbrace{0,\ldots 0}^{N},0,-1,0,1) \,.
\end{split}
\label{eq:phi_vev}
\end{align}
In this setup, we have a slightly different mass configuration as opposed to the AHPS setup.
While the fields $\mathcal{\chi}_{ab}$ and $\mathcal{\chi}_{N+i, a}$ are unchanged, the fields from the $M \times M$ part $\mathcal{\chi}_{N+i, N+i+1}$ now have masses $\sqrt{m_{i}^{2} + m_{i+1}^{2}}$ and {fields $\mathcal{\chi}_{N+i, N+j}$ have masses $m_{i}+m_{j}$}.
So, in the $m_{i} = m$ case, we have 
four types of fields: massless ones, fields of mass $m$, 
fields of mass $\sqrt{2} m$, and fields of mass $2m$.

\subsubsection*{Angle-dependent variant of \cite{Arkani-Hamed:2023epq} VEV pattern (AFHSTb)}\label{sec:angle_coulomb}

The above configuration can be generalized to include $SO(6)$ angles. 
This was mentioned in reference \cite{Correa:2012nk} in the context of Wilson loop appearing in Regge limits of Coulomb branch amplitudes.
For our purposes, we find it interesting to introduce two angles as follows,
\begin{align}\label{eq:AFHSTb}
\begin{split}
M_{\theta}=&\delta_{I8} \text{diag}(m_{1},m_{2} \sin(\theta_{24}),-m_{3}\cos(\theta_{13}),0) \\
&+ \delta_{I9} \text{diag}(0, -m_{2}\cos(\theta_{24}),-m_{3}\sin(\theta_{13}),m_{4}) \,.  
\end{split} 
\end{align}
Taking $\theta_{13}$ and $\theta_{24}$ to zero, this reduces to
\begin{align}
M_{I}=\delta_{I8} \text{diag}(m_{1},0,-m_{3},0) + \delta_{I9} \text{diag}(0, -m_{2},0,m_{4}) \,.
\end{align}
We recover the previous configuration \eqref{eq:vev_deformed} if we set $m_{1}=m_{3} = m_{x}$ and $m_{2}=m_{4} = m_{y}$. 

\subsection{Infrared-finite four-point amplitude}

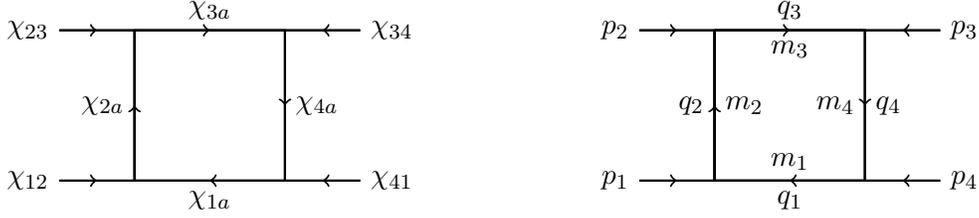
\begin{figure*}[t]
\hspace{1cm}
\begin{tikzpicture}
    \draw[thick] (1,0) node[left] {$\mathcal{\chi}_{12}$} -- (2,0);
    \draw[->-, thick] (1,0) -- (2,0)  {};
    \draw[thick] (1,2) node[left] {$\mathcal{\chi}_{23}$} -- (2,2);
    \draw[->-, thick] (1,2) -- (2,2)  {};
    \draw[thick] (4,0) -- (5,0) node[right] {$\mathcal{\chi}_{41}$};
    \draw[->-, thick] (5,0) -- (4,0)  {};
    \draw[thick] (4,2) -- (5,2) node[right] {$\mathcal{\chi}_{34}$};
    \draw[->-, thick] (5,2) -- (4,2)  {};
    %
    \draw[thick] (2,0) -- (2,2) node[midway,left] {$\mathcal{\chi}_{2a}$};
    \draw[->-, thick] (2,0) -- (2,2)  {};
    \draw[thick] (2,0) -- (4,0) node[midway,below] {$\mathcal{\chi}_{1a}$};
    \draw[->-, thick] (4,0) -- (2,0)  {};
    \draw[thick] (2,2) -- (4,2) node[midway,above] {$\mathcal{\chi}_{3a}$};
    \draw[->-, thick] (2,2) -- (4,2)  {};
    \draw[thick] (4,0) -- (4,2) node[midway,right] {$\mathcal{\chi}_{4a}$};
    \draw[->-, thick] (4,2) -- (4,0)  {};
    %
    \draw[thick] (2,0) -- (4,0) -- (4,2) -- (2,2) -- cycle;
   %
\end{tikzpicture}
\hspace{2cm}
\begin{tikzpicture}
    \draw[thick] (1,0) node[left] {$p_1$} -- (2,0);
    \draw[->-, thick] (1,0) -- (2,0)  {};
    \draw[thick] (1,2) node[left] {$p_2$} -- (2,2);
    \draw[->-, thick] (1,2) -- (2,2)  {};
    \draw[thick] (4,0) -- (5,0) node[right] {$p_4$};
    \draw[->-, thick] (5,0) -- (4,0)  {};
    \draw[thick] (4,2) -- (5,2) node[right] {$p_3$};
    \draw[->-, thick] (5,2) -- (4,2)  {};
    %
    \draw[thick] (2,0) -- (2,2) node[midway,left] {$q_{2}$};
    \draw[thick] (2,0) -- (2,2) node[midway,right] {$m_2$};
    \draw[->-, thick] (2,0) -- (2,2)  {};
    \draw[thick] (2,0) -- (4,0) node[midway,below] {$q_{1}$};
    \draw[thick] (2,0) -- (4,0) node[midway,above] {$m_1$};
    \draw[->-, thick] (4,0) -- (2,0)  {};
    \draw[thick] (2,2) -- (4,2) node[midway,above] {$q_{3}$};
    \draw[thick] (2,2) -- (4,2) node[midway,below] {$m_3$};
    \draw[->-, thick] (2,2) -- (4,2)  {};
    \draw[thick] (4,0) -- (4,2) node[midway,right] {$q_{4}$};
    \draw[thick] (4,0) -- (4,2) node[midway,left] {$m_4$};
    \draw[->-, thick] (4,2) -- (4,0)  {};
    %
    \draw[thick] (2,0) -- (4,0) -- (4,2) -- (2,2) -- cycle;
\end{tikzpicture}
\caption{
\label{fig:momentumkinematics_coulomb}
On the left: Field configuration on the Coulomb branch.
On the right: Momentum and mass configuration corresponding the four-point one-loop deformed amplituhedron. 
}
\end{figure*}

Let us consider in the framework of the discussed Coulomb branch configuration \eqref{eq:AFHSTb} the $2 \to 2$ scattering of `light' scalars from the $M\times M$ part of the matrices.
More precisely, we are interested in the following amplitude,
\begin{align}
A = \langle (\Phi_{I})_{N+1, N+2} (\Phi_{J})_{N+2, N+3}(\Phi_{I})_{N+3, N+4} (\Phi_{J})_{N+4,N+1} \rangle \,, 
\label{eq:coulomb_scalar_amplitude}
\end{align}
where $I,J \in \{4,\hdots, 7\}$ and $I \neq J$.
The tree level contribution is given by
\begin{align}
A^{\rm tree} = i g_{YM}^{2} \,.  
\end{align}
In what follows we will consider the normalized amplitude $M$, which is defined by
\begin{align}
A = A^{\rm tree}M \,.
\end{align}
It can be expanded in terms of the coupling constant as
\begin{align}\label{eq:m_amplitude}
M = 1 + g^{2} M^{(1)} + g^4 M^{(2)} + {\cal{O}}(g^6) \,,
\end{align}
where $g$ was defined in eq. (\ref{eq:defg}).
As was discussed in \cite{Alday:2009zm}, in the large $N$ limit, the massless fields from the unbroken part of the theory dominate in the loops.
They can interact with the chosen external states via the heavy fields from the off-diagonal part. The masses of the latter effectively provide an infrared regulator, making the amplitudes well-defined in four-dimensions.
We note that the mass configuration of this amplitude, see Fig.~\ref{fig:momentumkinematics_coulomb}, 
\begin{align}\label{eq:deformed_kinem}
&p_{i}^{2} = m_{i}^{2} + m_{i+1}^{2} \,, \\ 
&q_{i}^{2} = m_{i}^{2} \,,
\end{align}
matches that of the deformed Amplituhedron setup, as discussed in subsection \ref{subsec:Amplituhedron}. 
In the angle dependent case the we still have $q_{i}^{2} = m_{i}^{2}$ but now the external fields satisfy the following on-shell condition
\begin{align}
p_{i}^{2} = (m_{i+1}\vec{n}_{i+1}-m_{i}\vec{n}_{i})^{2} \,,   
\end{align}
where $\vec{n}_{i}$ vectors satisfy
\begin{align}\label{eq:n_vectors_products}
&\vec{n}_{i}^{2} = 1 \text{ for } i=1,\cdots ,4 
,, \notag \\
&\vec{n}_{1} \cdot \vec{n}_{2} = \sin (\theta_{24}) \,, \vec{n}_{1} \cdot \vec{n}_{3} = -\cos (\theta_{13}) \,, \vec{n}_{1} \cdot \vec{n}_{4} = 0 \,, \\
&\vec{n}_{2} \cdot \vec{n}_{3} = \sin (\theta_{13}-\theta_{24}) \,, \vec{n}_{2} \cdot \vec{n}_{4} = -\cos (\theta_{24}) \,, \vec{n}_{3} \cdot \vec{n}_{4} = -\sin (\theta_{13}) \,. \notag
\end{align}

The amplitude can be expressed as a function of the following two invariants
\begin{align}
\begin{split}
u_{\theta_{13}} = \frac{\left(-s+m_{1}^{2}+m_{3}^{2} + 2m_{1}m_{3}\cos(\theta_{13}) \right)^{2}}{4m_{1}^{2}m_{3}^{2}}\,,  \\
v_{\theta_{24}}= \frac{\left(-t+m_{2}^{2}+m_{4}^{2} + 2m_{2}m_{4}\cos(\theta_{24}) \right)^{2}}{4m_{2}^{2}m_{4}^{2}}  \,.
\end{split}
\end{align}

{To find the integrand of the one-loop contribution, we performed a direct calculation similar to Appendix B of \cite{Alday:2009zm}. We found that only the box diagram contributes. This confirms that the one-loop contribution has the expected dual conformal symmetry. Also, it matches the Amplituhedron expectation of eq. (\ref{eq:oneloopboxmomentum}), with the normalization given by 
\begin{align}
n^{(1)} = -\frac{1}{4} \,.
\label{eq:one_loop_normalization}
\end{align}
{
Introducing the notation
\begin{align}
q_{ij} = 1+ \frac{2 m_i m_j}{-x^2_{ij}+(m_1+m_j)^2}(\cos(\theta_{ij})-1) \,,\quad x_{13}^2 =s \,,\quad x_{24}^2 =t \,,
\end{align}
we find for the angle-dependent case 
\begin{align}
n^{(1)} = 
-\frac{1}{4} q_{13} q_{24}
\,.
\end{align}
}
{At two loops,
assuming eq. (\ref{eq:expectedtwoloopanswer}), we used a maximal unitarity cut \cite{Britto:2004nc} to compute the normalization, with the result
\begin{align}
n^{(2a)}= -\frac{1}{8} \,,
\end{align}
and in the angle-dependent case
\begin{align}
n^{(2a)} = 
-\frac{1}{8} q_{13}^2 q_{24} 
\,.
\end{align}
In the next sections we evaluate the two-loop amplitudes, and study their Regge limit.

%
%
%
%
%
%

\section{Finite loop integrals from differential equations}
\label{sec:de}

The one and two-loop level contribution to the amplitude \eqref{eq:m_amplitude} can be calculated by the method of the differential equations \cite{Kotikov:1990kg,Bern:1993kr,Gehrmann:1999as,Henn:2014qga}. (For reviews, see \cite{Argeri:2007up,Henn:2014qga}).
We know that the studied amplitude is both ultraviolet and infrared finite hence we can focus on the differential equations in four dimensions. 
This will also allow us to profit from the underlying dual conformal symmetry of the functions. The latter can be made manifest in the embedding formalism \cite{Dirac:1936fq,Weinberg:2010fx,Simmons-Duffin:2012juh}. As discussed in \cite{Caron-Huot:2014lda}, employing this formalism makes it easy to restrict ourselves to  integration by parts identities \cite{Chetyrkin:1981qh} that involve dual conformal integrals only, a welcome simplification.
In this section, we therefore follow closely the method described in \cite{Caron-Huot:2014lda}, applying it to the different kinematic setup of the Feynman integrals considered here.

\subsection{Embedding formalism for Feynman integrals}
\label{subsec:embedding_formalism}

The idea behind the embedding formalism is to embed four-dimensional space-time into five-dimensional projective space $\mathbb{CP}^{5}$ as
\begin{equation}
x^{\mu} \rightarrow X^{a} = (X^{\mu}, X^{-},X^{+})^{T} \,, \text{ with } X^{a} \simeq \alpha X^{a} \, (\alpha \neq 0) \,, 
\end{equation}
equipped with the following inner product, 
\begin{equation}
(XY) = 2 X^{\mu}Y_{\mu} + X^{-}Y^{+} + X^{+}Y^{-} \,.
\label{inner_product}
\end{equation}
Assigning a vector $Y$ as an integration variable allows us to represent Feynman propagators in terms of the inner product \eqref{inner_product}. 
For example, the four-particle kinematics discussed above can be realized as follows, 
\begin{align}\label{eq:x_embedding}
\begin{split}
&X_{1}^{a} =     \left(
\begin{matrix}
0^\mu \\
m_{1}^2\\
1
\end{matrix}
\right)  \,,\quad 
 %
 X_{3}^{a} =     \left(
\begin{matrix}
- (p_{1}+p_{2})^\mu \\
-(p_{1}+p_{2})^2+m_{3}^2\\
1
\end{matrix}
\right) 
\,,\\
 %
&X_{2}^{a} =     \left(
\begin{matrix}
- p_{1}^\mu \\
-p_{1}^2+m_{2}^2\\
1
\end{matrix}
\right) 
 \,, \quad
 %
  X_{4}^{a} =     \left(
\begin{matrix}
p_{4}^\mu \\
-p_{4}^2+m_{4}^2\\
1
\end{matrix}
\right) \,.
\end{split}
\end{align}
Imposing 
\begin{align}
(X_1 X_2) = (X_2 X_3) = (X_3 X_4) = (X_4 X_1) = 0
\end{align}
implies
\begin{align}
    p_{1}^2 = m_1^2 + m_2^2 \,, \quad 
        p_{2}^2 = m_2^2 + m_3^2 \,,\quad
            p_{3}^2 = m_3^2 + m_4^2 \,, \quad 
                p_{4}^2 = m_4^2 + m_1^2 \,,
\end{align}
which corresponds exactly to the kinematics \eqref{eq:deformed_kinem}.
Moreover, there are two dual conformal invariant variables in this kinematics, namely
\begin{align}\label{eq:cross_ratios}
\begin{split}
u=\frac{({X}_{1}{X}_{3})^{2}}{{X}_{1}^{2}{X}_{3}^{2}}  \,, \qquad 
v=\frac{({X}_{2}{X}_{4})^{2}}{{X}_{2}^{2}{X}_{4}^{2}}   \,.
\end{split}
\end{align}
One may check that these equations agree with eq. (\ref{eq:defdualconformalinvariantsuv}). The loop momentum $k$ appears in embedding space in the combination
\begin{align}\label{eq:y_embedding}
 Y^{a} =     \left(
\begin{matrix}
k^\mu \\
-k^2\\
1
\end{matrix}
\right) \,.
\end{align}
A quick calculation shows that with this choice, we have 
\begin{align}
({X}_1 Y) =& {-k^2 + m_{1}^2} \,,
\end{align}
and similarly for $({X}_i Y)$.
This allows us to translate the propagators of the box integral of eq. (\ref{eq:oneloopboxmomentum}) to the embedding space. The most basic integral (an analog of a massive tadpole) is the following,
\begin{align} 
\int_Y \frac{1}{(YQ)^4} =  \frac{1}{\Gamma(4)} \frac{1}{[\frac{1}{2} (QQ)]^2 } \,.
\label{measuredef}
\end{align}
Here the normalization is a choice, implicit in $\int_Y$, and we used the abbreviation
\begin{align}
\int_Y \equiv \int \frac{\delta{(\frac{1}{2}}(YY)) d^{4+2}Y}{\text{vol}(GL(1))} \,.    
\end{align}
Let us explain the integration measure and contour.
The notion $GL(1)$ corresponds to the projective invariance $Y \simeq \alpha Y$ and $\text{vol}(GL(1))$ indicates that we mod out by this freedom. 
A particular choice of $\text{vol}(GL(1))$ gauge fixing term is $\delta((Y W) -1)$, for some $W$.

The most general dual conformal one-loop Feynman integral we require is written in the embedding formalism as follows, 
\begin{equation}\label{eq:dualconformaloneloopintegral}
G_{a_{1}a_{2}a_{3}a_{4}} = \int \frac{d^{4+2}Y  \delta{(\frac{1}{2}}(YY))}{\text{vol}(GL(1))} \frac{1
}{({X}_1 Y)^{a_{1}} ({X}_2 Y)^{a_{2}}  ({X}_3 Y)^{a_{3}} ({X}_4 Y)^{a_{4}}} \, ,   
\end{equation}
where $\sum_{i=1}^{4} a_i = 4$ corresponds to the dual conformal covariance.
One can see that in the embedding formalism the propagators become linear in the integration variables, i.e., they represent hyperplanes, 
and the quadratic dependence is transferred into the integration measure in form of the $\delta(\frac{1}{2}YY)$ function. 

In practice, one can use eq. (\ref{measuredef}) in combination with the Feynman trick of combining propagators,
\begin{align}
\frac{1}{\prod_{i=1}^{n} A_i^{a_i}} = \frac{\Gamma(a_1 + \ldots a_n)}{\Gamma(a_1) \ldots \Gamma(a_n)}  \int_0^\infty \frac{\prod_{i=1}^{n} d\alpha_i \alpha_i^{a_i-1} }{ {\rm GL}(1)} \frac{1}{(\alpha_{1} A_1 + \ldots \alpha_{n} A_n)^{a_1 + \ldots a_n}} \,,
\end{align}
in order to derive a parametric integral representation for integrals of the type (\ref{eq:dualconformaloneloopintegral}).

\subsection{Integration by parts relations in the embedding space}

Integration by parts relations (IBP) can be generated directly in the embedding space in that way one can easily restrict to the subset of relations between dual conformal integrals in four dimensions only. 
To do so, we write 
\begin{equation}
0 = \int_{Y} \frac{\partial}{\partial Y^{a}} \left[ \delta\left( \frac{1}{2} (YY) \right) Q^{a}(Y) \right] \, ,  
\label{eq:ibp_relation}
\end{equation}
where $Q(Y)$ is homogeneous vector, with $Q^{a}(Y) = \alpha^{4-1}Q^{a}(\alpha Y)$. 
This ensures that eq. (\ref{eq:ibp_relation}) is invariant under $GL(1)$ transformations.
When using eq. (\ref{eq:ibp_relation}), we would like to avoid derivatives of $\delta(\frac{1}{2}(YY))$ function, since it has no interpretation in terms of Feynman integrals. 
This can be achieved by imposing the orthogonality condition 
\begin{equation}
Y_{a}Q^{a}(Y) = 0 \, .    
\end{equation}
Examples of admissible IBP vectors are
\begin{align}\label{one_loop_ibp_vec}
Q_{ij }^a \equiv  (Y {X}_j) {X}_i^a -   (Y {X}_i) {X}_j^a \, , ~~~   
\end{align}
for $(i,j) \in \{ 1,2,3,4 \}$.
It turns out that these are sufficient for our purposes at one-loop, while at two-loop we introduce further IBP vectors.
Using eq. \eqref{eq:ibp_relation} with the above vectors $Q_{ij}$ on the one-loop integrals, we get the following identities,
\begin{align} \label{eq:oneloopIBP}
\begin{split}
0 &= \int_{Y} \partial_{Y, a} \frac{ Q_{ij}^a}{({X}_1 Y)^{a_1} ({X}_2 Y)^{a_2} ({X}_3 Y)^{a_3} ({X}_4 Y)^{a_4}} \\
&= \sum_{k=1}^4 a_k (({X}_k {X}_i)   A_{j}^-   - ({X}_k {X}_j)  A_{i}^- ) A_k^+ G_{a_1, a_2, a_3, a_4}\, , 
\end{split}
\end{align}
where $A_i^{\pm}$ maps $a_i \rightarrow a_i \pm 1$ when acting on $G_{a_1, a_2, a_3, a_4}$. {For example, applying the IBP vector $Q^{a}_{13}$ to the integral $G_{1,2,1,0}$, we obtain \begin{align}\label{eq:one_loop_ibp_example}
&G_{1,3,1,-1} = \frac{1+y^2}{2y}G_{1,2,1,0} \,.
\end{align}
}
We will use this relation later.

\subsection{Derivative with respect to external parameters}

We would like to derive differential equations for embedding space integrals of the type (\ref{eq:dualconformaloneloopintegral}).
As was mentioned above, these are dual conformally covariant.
Strictly speaking, these functions depend on all non-zero scalar products $(X_i X_j)$ that we can form from in our kinematics, of which there are six.
However, the functions we are ultimately interested in are dual conformally invariant, 
and as such depend on the two parameters $u$ and $v$ only, cf. eq. (\ref{eq:defdualconformalinvariantsuv}).
We could choose an appropriate normalization factor in $X_i$ to make the integrals invariant under dual conformal transformations, but this would lead to a bulkier notation.
So here we prefer to use the dual conformal invariance in a different way, namely by making the following particular dual conformal gauge choice,
\begin{align}
\begin{split}
&({X}_{1}{X}_{1}) = ({X}_{3}{X}_{3}) = 2x \,, \quad ({X}_{2}{X}_{2}) = ({X}_{4}{X}_{4}) = 2y \,,  \\  
&({X}_{1}{X}_{3}) = 1 + x^2 \,, \quad
  ({X}_{2}{X}_{4}) = 1+y^2 \,, \quad ({X}_{i}{X}_{i+1}) = 0 \,.
\label{eq:x_y_gauge}
\end{split}
\end{align}
Here we introduced new parameters $x$ and $y$, which are related to $u$ and $v$ by
\begin{align}\label{eq:cross_ratios_x_y}
u= \frac{1}{4}\left( x + \frac{1}{x} \right)^{2} \, , \quad v = \frac{1}{4} \left(y +\frac{1}{y} \right)^{2} \,.     
\end{align}
In the following we assume this choice, and will therefore regard all embedding space integrals $G$ as functions of $x,y$ only. 

Our goal is then to find differential operators with respect to $x$ and $y$.
To do so, we define basis differential operators 
\begin{align}
O_{i,j} = ({X}_i   \partial_{{X}_j}) \,,
\end{align}
and use the chain rule to express these as derivatives w.r.t. $x,y$.
There is a subtlety that we need to be careful about.
We need to ensure that the differential operators we use are compatible with our gauge choice, eq.  \eqref{eq:x_y_gauge}. 
As an example for the $x$ derivative we require
\begin{align}
\partial_{x} ({X}_{1} {X}_{3}) &= \partial_{x} (1+x^{2}) = 2x \, , \\
\partial_{x} ({X}_{1}^{2}) &= \partial_{x} ({X}_{3}^{2}) = \partial_{x} (2x) = 2 \, , \\
\partial_{x} ({X}_{2} {X}_{4}) &= \partial_{x} ({X}_{2}^{2}) = \partial_{x} ({X}_{4}^{2}) = 0 \, .
\end{align}
We find the following differential operator satisfies all criteria,
\begin{align} \label{derivop}
\partial_x = \frac{1}{(-1+x)(1+x)} (x O_{1, 1} - O_{1, 3} - O_{3, 1} + 
x O_{3, 3}) \,.
\end{align}
An analogous definition holds for $\partial_y$.

\subsection{One-loop differential equations in four dimensions} \label{subsec:one_loop_diff_eq}

\begin{figure}[t]
\hspace{0.4cm}
\begin{tikzpicture}
\node at (1,4.05) {4};
\draw[thick] (1,3) ellipse (0.6cm and 0.8cm);
\draw[thick] (0.5,1.8) -- (1,2.2);
\draw[thick] (0.8,1.8) -- (1,2.2);
\draw[thick] (1.2,1.8) -- (1,2.2);
\draw[thick] (1.5,1.8) -- (1,2.2);
\node at (1,1) {$g_{1}$};
\end{tikzpicture}
\hspace{0.8cm}
\begin{tikzpicture}
    \draw[thick] (1.8,0) node[left] {} -- (3,0.5);
    \draw[thick] (2.1,2) node[left] {} -- (2.4,1.5);
    \draw[thick] (3,0.5) -- (4.2,0) node[right] {};
    \draw[thick] (3.6,1.5) -- (3.9,2) node[right] {};
    %
    \draw[thick] (3,0.5) -- (2.4,1.5) node[midway,left] {1};
    \node at (3,-0.7) {$g_{2}$};
    \draw[thick] (2.4,1.5) -- (3.6,1.5) node[midway,above] {2};
    \draw[thick] (3,0.5) -- (3.6,1.5) node[midway,right] {1};
\end{tikzpicture}
\hspace{0.8cm}
\begin{tikzpicture}
    \draw[thick] (1,0) -- (1.6,0.4);
    \node at (2.1,-0.5) {$g_{3}$};
    \draw[thick] (1,2) node[left] {} -- (1.6,1.6);
    \draw[thick] (2.5,1) -- (3.2,2);
    \draw[thick] (2.5,1) -- (3.2,0);
    %
    \draw[thick] (1.6,0.4) -- (1.6,1.6) node[midway,left] {2};
    \draw[thick] (1.6,0.4) -- (2.5,1) node[midway,below] {1};
    \draw[thick] (1.6,1.6) -- (2.5,1) node[midway,above] {1};
\end{tikzpicture}
\hspace{0.8cm}
\begin{tikzpicture}
    \draw[thick] (1,0) node[left] {} -- (2,0);
    \draw[thick] (1,2) node[left] {} -- (2,2);
    \draw[thick] (4,0) -- (5,0) node[right] {};
    \draw[thick] (4,2) -- (5,2) node[right] {};
    %
    \draw[thick] (2,0) -- (2,2) node[midway,left] {1};
    \draw[thick] (2,0) -- (4,0) node[midway,below] {1};
    \node at (3,-0.7) {$g_{4}$};
    \draw[thick] (2,2) -- (4,2) node[midway,above] {1};
    \draw[thick] (4,0) -- (4,2) node[midway,right] {1};
\end{tikzpicture}
\caption{
Diagrams corresponding to the function basis of the differential equation for the one-loop four point Coulomb branch amplitude. Numbers indicate powers of propagators.  
}
\label{fig:one_loop_masters}
\end{figure}
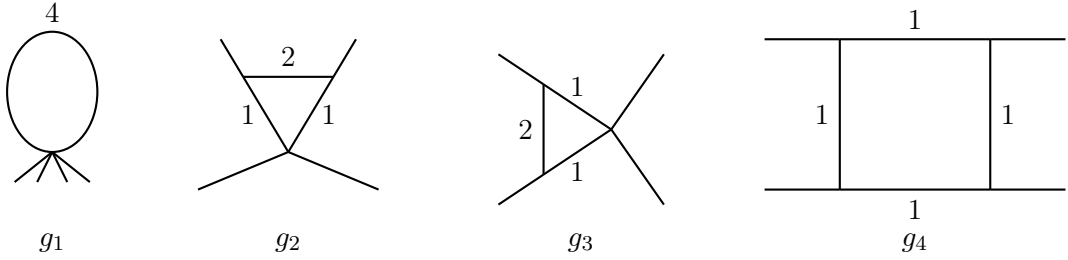

At one loop, we need to compute the box integral $G_{1,1,1,1}$, cf. eq. (\ref{eq:dualconformaloneloopintegral}) and Fig.~\ref{fig:one_loop_masters}.
As we will, see, only integrals of the type (\ref{eq:dualconformaloneloopintegral}) will be needed for this calculation.

By applying the differential operators $\partial_{x}$ and $\partial_{y}$ of eq. \eqref{derivop} to the box integral, we get
\begin{equation}  
\begin{split}
\partial_x G_{1,1,1,1} &= \frac{2x  G_{1,1,1,1}  - 2 G_{0,1,2,1} }{1-x^2}  \, ,  \\
\partial_y G_{1,1,1,1} &=  \frac{2 y  G_{1,1,1,1}   -2  G_{1,2,1,0} }{1-y^2}   \, ,
\end{split}    
\label{one_loop_de}
\end{equation}
where we used the symmetry $G_{2,1,0,1}=G_{0,1,2,1}$ (valid in the gauge of eq. (\ref{eq:x_y_gauge}), and after integration).
Let us first consider the homogeneous part of these equations,
\begin{equation}
\begin{split}
\partial_{x} f(x,y) = \frac{2x}{1-x^{2}}f(x,y) \,,\\
\partial_{y} f(x,y) = \frac{2y}{1-y^{2}}f(x,y) \,.
\end{split}
\end{equation}
The first of these equations has the solution
\begin{equation}
f(x,y) = \frac{1}{1-x^{2}}g(y) \, .
\label{f_homogeneous}
\end{equation}
Plugging this back into the second equation, the latter becomes
\begin{equation}
\partial_{y} g(y) = \frac{2y}{1-y^{2}}g(y) \,,
\end{equation}
with the solution
\begin{equation}
g(y) = \frac{c}{1-y^{2}} \,,   
\label{g_homogeneous}
\end{equation}
for some integration constant $c$.
Combining \eqref{f_homogeneous} and \eqref{g_homogeneous}, we have
\begin{equation}
f(x,y)=\frac{c}{(1-x^{2})(1-y^{2})} \,.
\end{equation}
This suggests introducing the normalized integral $g_{4}$
\begin{equation} \label{boxnorm}
G_{1,1,1,1} = \frac{1}{(1-x^{2})(1-y^{2})}g_{4} \,.
\end{equation}
Switching to this integral, eq. \eqref{one_loop_de} simplifies to as
\begin{equation}
\begin{split}
\partial_{x} g_{4} = -2(1-y^{2})G_{0,1,2,1} \, , \\
\partial_{y} g_{4} = -2(1-x^{2})G_{1,2,1,0} \, .
\end{split}    
\end{equation}
We remark that the choice of normalization factor for $G_{1,1,1,1}$ can also be found in an independent way, namely by computing the leading singularity of this integral, which is the residue evaluated at a global pole of the integrand \cite{Cachazo:2008vp}. Indeed, one can verify that the leading singularity of this integral is given by (again, in our gauge choice) $1/(1-x^2)(1-y^2)$. 
This normalization makes $g_4$ a pure function \cite{Arkani-Hamed:2010pyv}, and such functions are expected to satisfy canonical differential equations \cite{Henn:2013pwa}, in agreement with our calculation.

{To complete the system we iterate the procedure above with the triangle integrals $G_{0,1,2,1}$ and $G_{1,2,1,0}$ (cf. Fig.~\ref{fig:one_loop_masters}). (See also \cite{Caron-Huot:2014lda} for a more detailed discussion.)
We continue taking derivatives of the inhomogeneous parts, and consider IBP relations \eqref{eq:oneloopIBP}. 
One also needs to take into account the symmetry relation $G_{a_{1},a_{2},a_{3},a_{4}} = G_{a_{1},a_{4},a_{3},a_{2}}$ in the $s$-channel and similarly in the $t$-channel. 
After taking two derivatives, the leftover integral is $G_{0,0,0,4}$ whose derivative after IBP reduction has no inhomogeneous term which means that system closes and can be solved.}

{Let us illustrate this. We proceed by acting with the differential operators \eqref{derivop} on $G_{1,2,1,0}$,
\begin{align}
&\partial_{x} G_{1,2,1,0} = -\frac{G_{0,2,2,0}-2xG_{1,2,1,0}+G_{2,2,0,0}}{1-x^2} \,,\\
&\partial_{y} G_{1,2,1,0} = -\frac{2G_{1,3,1,-1}-2yG_{1,2,1,0}}{1-y^{2}} \,.
\end{align}
These equations can be further simplified with the help of eq.  \eqref{eq:one_loop_ibp_example}, and also using $G_{0,2,2,0}  = G_{2,2,0,0}$ and $G_{2,2,0,0} = \frac{3y}{x}G_{0,0,0,4}$, to get
\begin{align}
&\partial_{x} G_{1,2,1,0} = - \frac{6yG_{0,0,0,4}-2x^{2}G_{1,2,1,0}}{x(1-x^2)}\,,\\
&\partial_{y} G_{1,2,1,0} = -\frac{G_{1,2,1,0}}{y} \,.    
\end{align}
The second equation is already in a sought after form. To close the system we need to take derivatives of $G_{0,0,0,4}$,
\begin{align}
&\partial_{x} G_{0,0,0,4} = 0 \,, \\
&\partial_{y} G_{0,0,0,4} = -\frac{4G_{0,-1,0,5}-4yG_{0,0,0,4}}{1-y^2} \,.
\end{align}
The first equation already tells us that $G_{0,0,0,4}$ is a function only of the $y$ variable. The second equation can be further simplified by IBP relations \eqref{eq:oneloopIBP}
\begin{align}
&\partial_{y} G_{0,0,0,4} = -\frac{2G_{0,0,0,4}}{y} \,.
\end{align}
}
As a result of this procedure, we find a closed system of differential equations which involves four basis integrals $G_{1,1,1,1}$, $G_{0,1,2,1}$, $G_{1,2,1,0}$, and {$G_{0,0,0,4}$} (see Fig~\ref{fig:one_loop_masters} for the corresponding diagrams).
As explained above, we normalize them to be pure functions, as follows
{
\begin{align}\label{eq:one_loop_basis}
\begin{split}
    g_1 &= -24 y^2 \, G_{0,0,0,4} \, , \\
    g_2 &= 4 x (1-y^2) \, G_{0,1,2,1} \, , \\
    g_3 &= 4 (1-x^2) y \, G_{1,2,1,0} \, , \\
    g_4 &= -2 (1-x^2) (1-y^2) \, G_{1,1,1,1} \,  .
\end{split}
\end{align}
}
In terms of these functions, the $x$ and $y$ derivatives of $\vec{g} = (g_1, g_2, g_3, g_4)$ take the following canonical form \cite{Henn:2013pwa,Caron-Huot:2014lda},
\begin{align}
    \partial_x \vec{g} =  A_x \vec{g} \, , ~~~ \text{and} ~~~     \partial_y \vec{g} =  A_y \vec{g} \,,
\end{align}
where 
\begin{align}
 A_x = 
    \begin{pmatrix}
    0 & 0 & 0 & 0 \\
    0 & 0 & 0 & 0 \\
    \frac{1}{x} & 0 & 0 & 0 \\
    0 & \frac{1}{ x} & 0 & 0
    \end{pmatrix}
    \, , ~~~ \text{and} ~~~    A_y = 
    \begin{pmatrix}
    0 & 0 & 0 & 0 \\
    \frac{1}{y} & 0 & 0 & 0 \\
    0 & 0 & 0 & 0\\
    0 & 0 & \frac{1}{y} & 0
    \end{pmatrix} \,.
\end{align}
Introducing $d= dx \partial_x + dy \partial_y$, we can write these equations in a combined way as follows,
\begin{align}
    d \vec{g} = (d \tilde{A}) \vec{g} \, , 
\end{align}
where 
\begin{align}
        \tilde{A} = 
    \begin{pmatrix}
    0 &  0 & 0 & 0 \\
    \log\left( {y} \right) & 0 & 0 & 0 \\
    \log\left( {x} \right) & 0 & 0 & 0\\
    0 & \log\left( {x} \right) & \log\left( {y} \right) & 0
    \end{pmatrix} \,.
\end{align}
One may verify that $\partial_x \tilde{A} = A_x$ and $\partial_y \tilde{A} = A_y$. 
The boundary constants can be fixed to zero by requiring the solutions to be regular at $x=y=1$, and by computing the trivial tadpole integral. This gives
\begin{align}
\vec{g}(x=y=1) = (-4, 0,0,0)^{T} \,.
\end{align}
Taking this into account,
the solution to the differential equations is given by,
{
\begin{equation}\label{eq:one_loop_functions}
\begin{split}
&g_{1} = -4   \,, \quad 
g_{2} = -4 \log(y)  \,, \quad 
g_{3} = -4 \log(x) \,, \quad 
g_{4} = -4 \log(x) \log(y) \,. 
\end{split}
\end{equation}
}
This agrees with the direct calculation of the function $G_{1,1,1,1}$ by means of a Feynman parametric representation.

\subsection{Two-loop differential equations \label{subsec:twoloopDE}}

In this section, we discuss the differential equation system governing the double box integral needed for the two-loop amplitude. 
Let us defined the following two-loop integral family,
\begin{align} 
{G}_{a_1,a_2,a_3,a_4,b_1, b_2, b_3, b_4, c} := 
\int_{Y_1, Y_2}  \frac{1}{(Y_1 Y_2)^{c}}
\prod_{i=1}^{4} \frac{1}{({X}_i Y_1)^{a_i}  ({X}_i Y_2)^{b_i} } \,, 
\label{deftwoloopfamily2}
\end{align}
where $a_{1}+a_{2}+a_{3}+a_{4}+c=4$ and $b_{1}+b_{2}+b_{3}+b_{4}+c=4$, due to dual conformal symmetry. 
The $s$-channel double box integral can be represented by $G_{1,1,1,0,1,0,1,1,1}$. 
In this case $a_{4}\leq 0$ and $b_{2}\leq 0$ represent possible irreducible numerators that may arise when considering derivatives.
{
Similarly, the $t$-channel double box can be represented by $G_{1,1,0,1,0,1,1,1,1}$.
}

At two loop loops there are a number of new features compared to the one loop: additional IBP vectors are required; there are IBP relations that relate two- and one-loop Feynman integrals. Let us discuss them in turn.

At the two-loop level there are two IBP vectors of the type $Q_{ij}$ \eqref{one_loop_ibp_vec}, i.e.,
\begin{align} \label{eq:twoloopqij}
Q_{ij,1 }^a \equiv  (Y_1 {X}_j) {X}_i^a -   (Y_1 {X}_i) {X}_j^a \, , ~~~  Q_{ij,2}^a \equiv  (Y_2 {X}_j) X_i^a -   (Y_2 {X}_i) {X}_j^a \, , 
\end{align}
for $(i,j) \in \{ 1,2,3,4 \}$, which we can use for deriving IBP relations in $Y_1$ and $Y_2$, respectively.
However, these turn out not to generate all necessary IBP relations for a complete reduction. 
Obviously, a crucial new element at two loops is the internal propagator $(Y_{1}Y_{2})^{-c}$, 
$c \in \mathbb{Z}$. It is natural to add the following IBP vectors,
\begin{align} 
Q_{i,2}^a \equiv  (Y_1 {X}_i) Y_{2}^{a} -   (Y_1 Y_{2}) {X}_i^a \, , ~~~  Q_{i,1}^a \equiv  (Y_2 {X}_i) Y_1^a -   (Y_2 Y_1) {X}_i^a \, . 
\label{eq:ibp_contact_term_vector}
\end{align}

An important subtlety has to do with taking derivatives of massless propagators, which can lead to contact terms in four dimensions, due to the Laplace equation. In our context, this can happen as follows,
\begin{align} \label{eq:loopreductionIBP}
\frac{\partial}{\partial Y_{1}^{a}} \frac{1}{(Y_{1}Y_{2})^{2}}Q_{i,2}^{a} = -2\delta^{(4)}(Y_{1},Y_{2}) \frac{(I Q_{i,2})}{(IY_{1})} + \text{regular terms} \, ,
\end{align}
and similarly for the $Y_{2}$ derivatives. As a consequence, the contact term will produce, through the $\delta$ function, one-loop integrals. 
Thus, we will generate IBP relations connecting two- and one-loop integrals. 

However, if one acts with the vectors \eqref{eq:ibp_contact_term_vector} on functions involving only a single power of the internal propagator $(Y_{1}Y_{2})^{-1}$ as a result one will obtain IBP relations involving factorized integrals (two-loop integrals where the power of the internal propagator is equal to zero, i.e., $(Y_{1}Y_{2})^{0}$).
For example by taking $Y_{1}$ derivative of $G_{0,1,2,0,0,0,2,2,1}$ with the IBP vector $Q^{a}_{2,2}$ of \eqref{eq:ibp_contact_term_vector} we get
\begin{align}
G_{0,0,3,0,0,0,1,2,1} = y \, G_{0,2,2,0,0,0,2,2,0} \,,   
\end{align}
which relates two-loop integral to the factorized integral.

To derive the differential equation system at the two-loop we use the same differential operators \eqref{derivop} as in the previous section.
In order to simplify the system, we use IBP relations with the IBP vectors \eqref{eq:twoloopqij}, 
\begin{equation}
    0 = \int_{Y_1, Y_2} \partial_{Y_k, a} 
    \left[ \frac{Q_{ij,k}^a}{(Y_1 Y_2)^{c}}
\prod_{w=1}^{4} \frac{1}{({X}_w Y_1)^{a_w}  ({X}_w Y_2)^{b_w} }  \right]\,,
\end{equation}
for $k = 1,2$. We also use symmetry relations between two-loop integrals which is a reflection of the propagators with respect to the middle propagator $(Y_{1}Y_{2})$ in the $s$-channel,
\begin{equation}
    G_{a_1, a_2, a_3, a_4, b_1, b_2, b_3, b_4, 1} -  G_{b_1, b_4, b_3, b_2, a_1, a_4, a_3, a_2, 1} = 0 \, . 
\end{equation}
For both of these types of relations, we are free to choose any integer value for the indices $a_1, \ldots, a_4, b_1, \ldots, b_4, c$ 
(except for keeping $a_{4}\leq 0$ and $b_{2}\leq 0$ to stay within the double box topology).
We find that choosing the indices $a_i$ and $b_i$ to be in the range $-1, \ldots , 2$, while fixing $c= 1$, 
is sufficient for simplifying the double box differential equations. 

Furthermore, we use loop reduction IBP relations, from the identity \eqref{eq:loopreductionIBP}, which introduces relations between two- and one-loop integrals
\begin{equation}\label{eq:ibp_one_two_reduction}
    0 = \int_{Y_1, Y_2} \partial_{Y_k, a} 
     \left[ \frac{Q_{i,k}^a}{(Y_1 Y_2)^{c}}
\prod_{w=1}^{4} \frac{1}{({X}_w Y_1)^{a_w}  ({X}_w Y_2)^{b_w} }  \right]\,,
    \end{equation}
for $k = 1,2$. 
Again, we keep $a_{4}\leq 0$ and $b_{2}\leq 0$, but the remaining indices $a_i$ and $b_i$ can take values in the range $-1, \ldots , 2$ (while we fix $c= 2$). 
Let us look at an example of this loop reduction IBP relation. By acting with $Y_{1}$ derivative on $G_{0, 2, 0, 0, 0, 0, 2, 1, 2}$ with the IBP vector $Q^{a}_{2,2}$ we get
\begin{align}\label{eq:ibp_one_two}
G_{0,3,0,0,0,0,2,1,1} = \frac{1}{2y}G_{0,1,2,1} \,,    
\end{align}
connecting two-loop integral to a one-loop triangle.

Finally, we also use IBP relations between one-loop integrals, i.e. \eqref{eq:oneloopIBP}, as discussed above, but this time for a larger range of indices $a_i$ to be in the range $-3, \ldots, 4$.
From the relation like \eqref{eq:ibp_one_two} and other obtained from the \eqref{eq:ibp_one_two_reduction} we can see that after certain iteration of derivatives we will end up with the one-loop differential system.

In order to fully describe the system of differential equation we need $11$ functions $g_{i}$ $i=1,\hdots, 11$. 
{The $g_{1}, \hdots g_{4}$ functions corresponds to the one-loop system of eq. \eqref{eq:one_loop_functions}, and the remaining pure functions are defined in the following way}
\begin{equation} \label{eq:2loopGs}
\begin{split}
&g_{5} = -2 x^{2} (1-y^{2})^{2} (G_{0,1,2,1})^{2} \, , \\
&g_{6} = 128 x y^{3} (G_{1,3,0,0})^{2} \, , \\
&g_{7} = -2 y^{2} (1 - x^{2})^{2} (G_{1,2,1,0})^{2} \, , \\
&g_{8} = 8 x y ( 1-x^{2}) \left[ G_{0,2,1,0,1,0,2,0,1}+x y (G_{1,2,1,0})^{2} \right] \, , \\
&g_{9} = 2 y (1-x^{2})^{2} G_{1,1,1,0,1,-1,1,2,1} \, , \\
&g_{10} = 4 x(1-x^{2})(1-y^{2}) G_{0,1,2,0,1,0,1,1,1} \, \\
&g_{11} = - (1-x^{2})^{2}(1-y^{2}) G_{1,1,1,0,1,0,1,1,1} \, ,  
\end{split}
\end{equation}
where the functions on the right-hand side belong either to the one-loop family \eqref{eq:dualconformaloneloopintegral} or to the two-loop family \eqref{deftwoloopfamily2}. 
The functions $g_1, \ldots, g_{11}$ are normalized 
such that they 
satisfy the following canonical system of differential equations
\begin{equation} \label{eq:canonicaldiffeq}
d \vec{g} =(d\, A) \vec{g} \,,
\end{equation}
where the non-zero elements of the $A$ matrix are given by
\begin{align}
\label{eq:diff_eq_mat_els}
\begin{split}
&A_{2,1} = \log(y) \,, \quad 
A_{3,1} = \log(x) \,, \quad 
A_{4,2} = \log(x) \,, \quad
A_{4,3} = \log(y) \,,  \\
&A_{5,2} = \log(y) \,, \quad 
A_{7,3} = \log(x) \,, \quad 
A_{8,3} = \log \left[\left(1-x^2\right)^2\right] \,,  \\
& A_{9,4} = \log\left[  \frac{x^{2}-y^{2}}{1-x^{2}y^{2}}\right] \,, \quad  
A_{9,5} = \log \left[ \frac{x^{2}(1-y^2)^{2}}{(1-x^{2}y^{2})(x^{2}-y^{2})} \right] \,,  \\
&A_{9,6} = \log \left[\frac{(x+y) (1+x y)}{x (1+y)^2}\right] \,, \quad 
A_{9,7} = \log \left[  \frac{y^{2}}{(1-x^{2}y^{2})(x^{2}-y^{2})}   \right] \,,  \\
&A_{9,8} = \log(x) \,, \quad 
A_{10,4} = \log \left[\frac{y^2(1-x^2)^2}{ (1-x^{2} y^{2}) (x^{2}- y^{2})}\right] \,,  \\
&A_{10,5} = \log \left(  \frac{x^{2}-y^{2}}{1-x^{2}y^{2}} \right) \,, \quad 
A_{10,6} = \log \left( - \frac{1+xy}{x+y} \right) \,,  \\
&A_{10,7} = \log \left[  \frac{x^{2}-y^{2}}{y^{2}(1-x^{2}y^{2})}
\right] \,, \quad 
A_{10,8} = \log(y) \, \quad 
A_{11,9} = \log(y) \,  \\
&A_{11,10} = \log(x) \,. 
\end{split}
\end{align}
As at one-loop, we note that going to $x=y=1$ we obtain a boundary value almost without any effort.
It reads
\begin{align}\label{eq:twoloopboundary}
\vec{g}(x=1,y=1) = (-4, 0, 0, 0, 0, 2\pi^2 , 0, 0, 0, 0, 0)^{T} \, .    
\end{align}
{The $\pi^{2}$ in the boundary value corresponds to $g_{6}$ integral which is a product of one-loop integrals that have not appeared before. Interestingly, the latter evaluate to $\pi$, as an elementary Feynman parameter evaluation shows.} 

We collect the vector of basis functions $\vec{g}$, $A$ matrix and the boundary vector in the ancillary file \texttt{differential\_equation\_ancillary.txt}.

\begin{figure}[t]
    \centering
    \includegraphics[width=\textwidth]{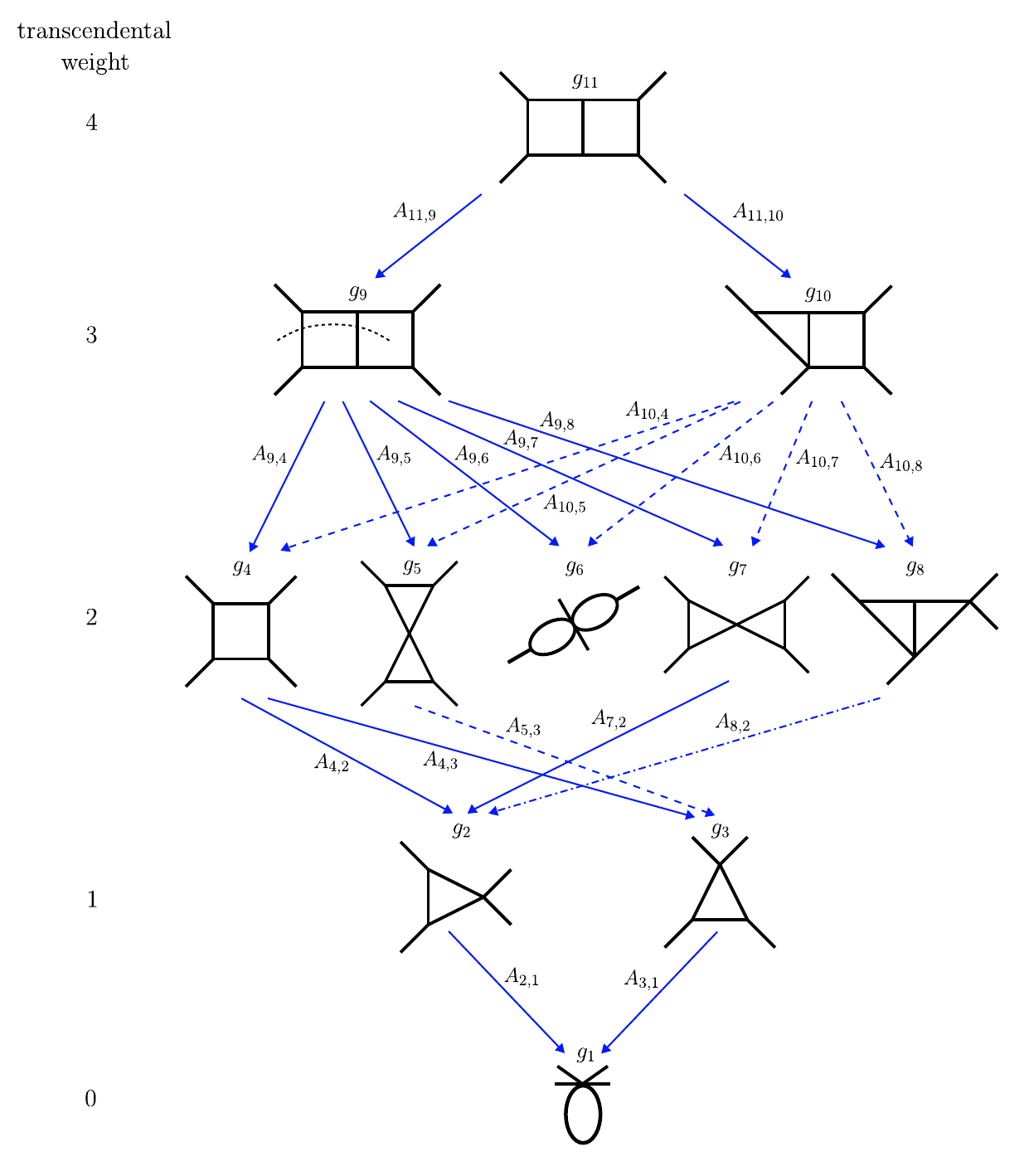}
    \caption{Differential equation tree. Arrows indicate how the functions are linked through $d \log$ derivatives (see \eqref{eq:canonicaldiffeq} and \eqref{eq:diff_eq_mat_els}). 
    }
    \label{fig:diffeqtree}
\end{figure}
The non-zero entries of the matrix $A$ are visualized in Fig.~\ref{fig:diffeqtree}. 
For example,  the derivative of $g_{11}$ with respect to $x$ is equal to $d \log(x) g_{10}$, which is represented in Fig.~\ref{fig:diffeqtree} by the arrow labeled with $A_{11,10}$.

As explained in \cite{Caron-Huot:2014lda}, Fig.~\ref{fig:diffeqtree}, allows one to immediately read off the solution to eq. (\ref{eq:canonicaldiffeq}), in terms of iterated integrals:
The answer for a given integral is given by summing over all branches connecting it to $g_1$, with the summands being iterated integrals, and with the integration kernels given by the non-zero matrix elements of the corresponding edges of the graph. Combining this insight with the boundary value (\ref{eq:twoloopboundary}) confirms that the basis functions $\vec{g}$ are indeed pure functions, with the weight as indicated in Fig.~\ref{fig:diffeqtree}.
As a byproduct, the symbol alphabet \cite{Goncharov:2010jf} can be read off from the differential equation matrix $A$, i.e. by collecting all independent arguments of the logarithms in eq. (\ref{eq:diff_eq_mat_els}). This yields
\begin{equation}\label{eq:symbolalphabet}
\mathcal{A} = \lbrace x, y, 1-x, 1+x, 1-y, 1+y, 1-xy, 1+xy, x-y,x+y \rbrace \, . 
\end{equation}
This information can be used to study further the relevant function space, as we discuss presently.

\subsection{Singularity structure, function space, and polylogarithmic result}

{The key information about the functions  $g_{1},\hdots,g_{11}$ from eqs. \eqref{eq:one_loop_basis}, \eqref{eq:2loopGs} is contained in the differential equation matrix $A$, cf. eq. (\ref{eq:diff_eq_mat_els}). 
In particular, all locations where an element of the symbol alphabet (\ref{eq:symbolalphabet}) vanishes or becomes infinite corresponds to a potential singularity of one of the functions. Note that the functions are multivalued in general.
The word `potential' means that a singularity might occur not on the first sheet of a function, but only after analytic continuation.

The information about the singularity structure of an integral could in principle also be obtained by a  Landau analysis \cite{Bjorken_phd,Landau:1959fi,nakanishi}. This is a topic of current interest, see e.g. \cite{Fevola:2023fzn,Caron-Huot:2024brh} and references therein. We note that in \cite{Hannesdottir:2024hke} the Coulomb-branch integrals of reference \cite{Caron-Huot:2014gia} were analyzed in this way.

We already mentioned that the solution to the DE system  \eqref{eq:canonicaldiffeq}, together with the boundary values (\ref{eq:twoloopboundary}), can be given in terms of iterated integrals.
Here we wish to provide another representation of the answer, in terms of more well-known functions. We refer interested readers to \cite{Caron-Huot:2014lda} for a discussion of advantages and disadvantages of different representations of the answer.

A good starting point is the result of \cite{Goncharov:2010jf}, which tells us what the class of polylogarithmic functions is that we need up to weight four,
namely the logarithm, ${\rm Li}_{k}$ for $k=2,3,4$, and finally ${\rm Li}_{2,2}$. 
It is useful to start by reading off the symbol of the answer from the iterated integrals.
Reference \cite{Goncharov:2010jf} also provides a test for whether ${\rm Li}_{2,2}$ functions are needed in the answer. Applying the projection operator of eq. (17) in that reference to the symbol of $g_{11}$, we find zero. This means that no ${\rm Li}_{2,2}$ is needed, a welcome simplification. Note that the same is not true in the AHPS setup \cite{Alday:2009zm}.

This means that we can limit ourselves to the classical polylogarithms. We  can construct a list of admissible function arguments for these (see e.g. \cite{Duhr:2012fh}).
These can be organized into symmetry orbits {of the group $S_{3}$. It is generated by two elements $\sigma_{2}$ and $\sigma_{3}$, that act on the rational functions as 
\begin{align}
\begin{split}
&\sigma_{2}(f) = 1- f \,, \qquad 
\sigma_{3}(f) = \frac{1}{f} \,.
\end{split}
\end{align}
}
{For example, acting with the group $S_{3}$ on the symbol letter $x$, we get the following orbit,
\begin{align}
S_{3}(x) = \bigg\{ x, 1-x, \frac{1}{1-x}, -\frac{x}{1-x}, -\frac{1-x}{x}, \frac{1}{x}  \bigg\} \,.   
\end{align}
}
Due to well-known identities between ${\rm Li}_{2}, {\rm Li}_{3}, {\rm Li}_{4}$ functions, it turns out that at most one, two, three arguments from each orbit are needed for those functions, respectively.
The `at most' is due to the fact that there may be further identities involving functions with arguments taken from different orbits. 
Thankfully, the symbol / differential method allows one to conveniently keep track of such identities. For pedagogical examples, see section 5 of ref. \cite{Henn:2020omi}.
The identities mean that there is considerable freedom in how to represent the answer in terms of polylogarithms.
Whenever possible, it is desirable to choose arguments that make the branch cut structure of the functions manifest.
In particular, for $x>0, y>0$, which corresponds to the so-called Euclidean region, all integrals $\vec{g}$ are real-valued and free of branch cuts.
So whenever possible, we choose function arguments that make this behavior manifest. 
{
For example, consider the $g_{8}$ function, which can be integrated to the following function,
\begin{align}
g_{8}(x) = -4 \, \text{Li}_2\left(x^2\right)-8 \log (x) \log \left(1-x^2\right)+\frac{2 \pi ^2}{3} \,.    
\end{align}
However, individual terms in this expressions contain branch cuts at $x=1$, which must be spurious.
 The procedure described above allows to write it in a form that is manifestly free from branch cuts in the Euclidean region.
 This can be achieved for all but one of the weight-four functions.
In this way we find, in addition to eqs. (\ref{eq:one_loop_functions}),
\begin{align} 
\begin{split}
&g_{5} = -2 \log^{2}(y) \, , \quad
g_{6} = 2 \pi^2 \, , \quad
g_{7} = -2 \log^{2}(x) \, , \quad
{g_{8} = 4 \, \text{Li}_2\left(1-x^2\right)} \,, 
\end{split}
\end{align}
and
\begin{align}
    \begin{split}
    &g_{9} = 4 \text{Li}_3\left(1-\frac{1}{x^2}\right)+4 \text{Li}_3\left(1-x^2\right)-2\text{Li}_3\left(1-\frac{1}{y^2}\right)+\text{Li}_3\left(1-\frac{1}{x^2   y^2}\right)  \\
&+\text{Li}_3\left(1-\frac{x^2}{y^2}\right) -2\text{Li}_3\left(1-y^2\right)+ \text{Li}_3\left(1-\frac{y^2}{x^2}\right)+   \text{Li}_3\left(1-x^2 y^2\right)  \\
&-4 \text{Li}_2\left(1-x^2\right) \log \left(x\right) - 2\text{Li}_2\left(1-\frac{x^2}{y^2}\right) \log\left(x \right) -  2\text{Li}_2\left(1-x^2 y^2\right) \log \left(x\right)  \\
&+ 2\text{Li}_2\left(1-\frac{x^2}{y^2}\right) \log \left(y\right) +4 \text{Li}_2\left(1-y^2\right) \log \left(y\right) -  2\text{Li}_2\left(1-x^2 y^2\right) \log \left(y\right)  \\
&-8 \log ^3\left(x\right) - 12 \log ^2\left(y\right) \log \left(x\right)+ 4\log ^3\left(y\right)   \\
&+2\pi ^2 \left[-2 \log \left( 1 + y\right)+\log \left(1 + \frac{y}{x}\right)+\log \left(1 + x y\right)\right] \, ,
\end{split}
\end{align}
and 
\begin{align}
\begin{split}
&g_{10} =  \text{Li}_3\left(1-x^2 y^2\right) -\text{Li}_3\left(1-\frac{y^2}{x^2}\right)-\text{Li}_3\left(1-\frac{x^2}{y^2}\right)+\text{Li}_3\left(1-\frac{1}{x^2 y^2}\right)  \\
&-2 \text{Li}_2\left(1-x^2 y^2\right) 
\log \left( x y \right)
+2 \text{Li}_2\left(1-\frac{x^2}{y^2}\right) 
\log\left( \frac{x}{y} \right)
+4 \text{Li}_2\left(1-x^2\right) \log (y) \\
&
-8 \log ^2(x) \log (y)-2 \pi ^2 \log \frac{x+y}{1+x y}  -4 \log ^3(y) 
\,.
\end{split}
\end{align}
The function $g_{11}$ can be represented in terms of polylogarithms \cite{Arkani-Hamed:2023epq} as follows,
\begin{align}\label{eqg11new}
    g_{11}(x,y) = Q(y^{2}) -\frac{1}{2} Q\left( \frac{y^{2}}{x^{2}} \right) -\frac{1}{2} Q(x^{2} y^{2} ) - J_3(x^{2}) \log (y) \,,
\end{align}
where
\begin{align}
\begin{split}
    Q(z) =& 3 J_{4}(z) + \frac{3 \pi^4}{10} +\frac{\pi^2}{4} \log^{2} (z) +\frac{1}{4} \log^{4} (z) + \log^{2}(z)  {\rm Li}_{2}(1-z)  \\ &+ 4 \pi^2 {\rm Li}_{2}(-\sqrt{z}) - \log (z)  {\rm Li}_{3}\left( 1- \frac{1}{z} \right) - \log (z) {\rm Li}_{3}\left( 1- {z} \right)  \,, 
\end{split}\\
\begin{split}\label{eq:j4}
J_4 (z) =& \text{Li}_4 (z) - \log(z) \text{Li}_3 (z) + \frac{1}{2} \log^2 (z) \text{Li}_2 (z) + \frac{1}{6} \log^3 (z) \log(1-z) - \frac{1}{48} \log^4 (z) \, ,
\end{split}\\
\begin{split}\label{eq:j3}
J_{3}(z) =& \frac{1}{4} \log^3 \left(z \right) + \log \left(z\right)  {\rm Li}_{2}\left( 1- {z} \right) -2  {\rm Li}_{3}\left( 1- {z} \right) -2  {\rm Li}_{3}\left( 1- \frac{1}{z} \right) \,.
\end{split}
\end{align}
The formulas presented above for $g_1, \ldots, g_{10}$ are manifestly real-valued for $x>0,y>0$. For $g_{11}$, the same is true only for $0<x<1,0<y<1$.

In order to verify our analytic results, we performed the following numerical check.
For example, using a loop-by-loop approach, one may obtain the following Feynman parametrization for $g_{11}$, 
\begin{equation}
\begin{split}
g_{11} =- 
\int \frac{d^{3} \alpha}{GL(1)}\int \frac{d^4 \gamma}{GL(1)} \frac{(1-x^{2})^2(1-y^{2})}{\Big(\alpha_{1}^{2}x +\alpha_{2}^{2}y+\alpha_{3}^{2}x +\alpha_{1}\alpha_{3}(1+x^{2})\Big)} \times \\
\frac{}{\Big(\gamma_{1}^{2}x + \gamma_{2}^{2}(\alpha_{1}^{2}x +\alpha_{2}^{2}y+\alpha_{3}^{2}x +\alpha_{1}\alpha_{3}(1+x^{2})) + \gamma_{3}^{2}x+\gamma_{4}^{2}y + \gamma_{1}\gamma_{3}(1+x^2)+} \\
\frac{}{\gamma_{1}\gamma_{2}(2\alpha_{1}x+\alpha_{3}(1+x^{2})) + \gamma_{2}\gamma_{3}(2 \alpha_{3}x +\alpha_{1}(1+x^2)) + \gamma_{2}\gamma_{4}\alpha_{2}(1+y^2)\Big)^{2}} \,.
\end{split} \label{eq:g11feynman}
\end{equation}
We evaluated this numerically, for several parameter pairs of $(x,y)$ within the Euclidean region, using  \texttt{NIntegrate} in \texttt{Mathematica}.
We found numerical agreement with the values obtained from eq. \eqref{eqg11new}, within the error bars reported  by  \texttt{Mathematica}.

\subsection{Results for full amplitude, and in the Regge limit}
\label{subsec:limits}

With the definitions made above, 
the one- and two-loop corrections to the amplitude in eq. (\ref{eq:m_amplitude}) are
\begin{equation}\label{eq:M1intermsofg}
M^{(1)} = -{\frac{1}{2}}\frac{(1-x) (1-y)}{(1+x) (1+y)}g_{4} \,,
\end{equation}
and
\begin{align}\label{eq:M2intermsofg}
M^{(2)} (x,y) =   -\frac{(1-x)^2 (1-y)}{(1+x)^2 (1+y)} g_{11}(x,y)    -\frac{(1-x) (1-y)^2}{(1+x) (1+y)^2} g_{11}(y,x)  \,.
\end{align}
Given the results of the last subsection, the amplitude is known analytically. 
It can also be straightforwardly evaluated numerically for $0<x, 0<y$.

In view of applications in section \ref{sec:Regge}, we also provide the formulas relevant for computing the Regge limit.
There are two ways of obtaining these: as our results are rather simple, one can use the basic definitions of the polylogarithms to perform the expansions (using, for example, appropriate \texttt{Mathematica} commands). 
Alternatively, one can use the canonical DE to determine this expansion, which allows one to systematically obtain power suppressed terms to any order. Let us briefly outline this procedure (more details can be found in references \cite{Henn:2013nsa,Bruser:2018jnc}).

In this approach, one starts with the boundary values at $(x,y)=(1,1)$. One then uses the differential equations \eqref{eq:canonicaldiffeq} to transport this value to other points in the $(x,y)$-plane. 
It is convenient to do so along special paths, so no unnecessarily integration constants are introduced. For example, integrating just along the $y$ direction, and matching with the singular behavior as $y$ approached $0$, one may obtain the regularized boundary value at $(x,y) = (1,0^{+})$. Next, one integrates along $x$ to obtain the regularized boundary value $\vec{g} (x, 0^{+} )$, which now has the full $x$ dependence.
With this information at hand, one can expand the solution around that boundary point, for small $y$ (and arbitrary $x$). In this way, it is straightforward to obtain the solution including power suppressed terms in $y$. 
This approach also makes it clear that the function space in $x$ in the limit consists of harmonic polylogarithms \cite{Remiddi:1999ew,Maitre:2005uu} only, i.e. functions corresponding to the alphabet $\{x,1-x,1+x \}$. 

Following this procedure, we find the following results.

The one-loop box is simply
\begin{align}\label{eq:Regge1loopbox}
g_{4}(x,y) \stackrel{{y \to 0}}{=} -4\log(x)\log(y) + \mathcal{O}(y^{2}) \,.    
\end{align}
For the double box in the $s$-channel and $t$-channel, we obtain, respectively, \begin{align}\label{eq:Regge2loopbox1}
 \begin{split}
g_{11}(x,y) \stackrel{{y \to 0}}{=}& 
  - \log^{2}(y) 2\log^{2}(x) \\
  & \hspace{-1.5cm} - 
  2\log(y)\bigg[
2\text{Li}_3(x^2)-2\log(x)\text{Li}_2(x^2)- 
 \frac{1}{3}(\pi^{2}+2\log^2(x))\log(x) - 2 \zeta_{3}
 \bigg] \\
 & \hspace{-1.5cm} - 
 \left[ \pi^2+\frac{1}{3}\log^{2}(x)\right]\log^{2}(x) 
  + y \frac{2 \pi^{2}(1-x)^{2}}{x} + \mathcal{O}(y^{2})\,. 
 \end{split}
 \end{align}
and
\begin{align}\label{eq:Regge2loopbox2}
 \begin{split}
 g_{11}(y,x) \stackrel{{y \to 0}}{=}& -\frac{4}{3} \log(y) \log(x) \left( \pi^{2} + \log^{2}(x) \right) \\
 & \hspace{-2cm} + \bigg[
 3 \text{Li}_4\left(x^2\right) -4 \text{Li}_3\left(x^2\right) \log (x)  +2 \text{Li}_2\left(x^2\right) \log ^2(x)+4 \pi ^2 \text{Li}_2(-x) \\
 &\hspace{-1.5 cm} + \frac{1}{3}{\log ^4(x)}  +\pi ^2 \log ^2(x) + 4 \zeta_{3} \log (x) +\frac{3 \pi ^4}{10} 
 \bigg]  - 2 \pi^2
 \frac{(1-x^2)}{x}y + \mathcal{O}(y^{2}) \,.
 \end{split}
 \end{align}
These formulas, including additionally also terms up to order $\mathcal{O}(y^{3})$ order, are collected in the ancillary file
\texttt{regge\_ancillary.txt}.

%
%
%
%
%
%

\section{Regge limit and anomalous dimensions}
\label{sec:Regge}

\subsection{$SO(4)$ invariant Regge limit}

It is interesting to study the Regge limit $(-t) \to \infty$ (at fixed $s$) of the scattering amplitude, which corresponds to $y \to 0$, keeping $x>0$ fixed. 
At leading power, one expects the amplitude to exponentiate in this limit, with the exponent being given by the (angle-dependent) cusp anomalous dimension.
In reference  \cite{Caron-Huot:2014gia,Bruser:2018jnc}, it was pointed out that the dual conformal symmetry enlarges the usual $SO(3)$ symmetry of partial wave expansions to $SO(4)$.
Using an $SO(4)$-invariant expansion parameter, it was noticed in reference \cite{Bruser:2018jnc} that the first power suppressed terms in that expansion also exponentiate, 
and that the exponent is given by a cusped Wilson loop with a scalar operator insertion.

We repeat the analysis of reference \cite{Bruser:2018jnc}, but for the kinematic setup considered here.
As a result, we find the following  $SO(4)$-invariant angle,
\begin{align}
\cos(\vartheta) = \frac{m_{2}^{2}+m_{4}^{2}-t}{2\sqrt{m_{2}^{2}m_{4}^{2}}} \,. 
\label{eq:cos_theta}
\end{align}
For large $-t$, this angle becomes imaginary and large, and satisfied the simple relation 
\begin{align}
y=e^{-i \vartheta} \,.
\end{align}
Repeating the partial wave analysis of ref. \cite{Bruser:2018jnc} for our setup suggests that the amplitude normalized in the following way,
\begin{align}\label{MnormalizedforRegge}
\frac{1+y}{1-y} M(x,y)\,,
\end{align}
should have a simple expansion in $y$.

In section \ref{sec:de}, we computed the perturbative results up to two loops, and to the first power suppressed terms in $y$, cf. eqs. (\ref{eq:M1intermsofg}) ,(\ref{eq:M2intermsofg}) and  (\ref{eq:Regge1loopbox}), (\ref{eq:Regge2loopbox1}), and (\ref{eq:Regge2loopbox2}).
Indeed, taking into account the perturbative results derived in the previous section, we find that, up to the subleading order in $y$, the amplitude has an expansion in terms of just two exponentials
\begin{align}
\lim_{y \to 0} \frac{1+y}{1-y} M(x,y) = r_{0}(x)y^{-j_{0}(x)-1} + r_{1}(x)y^{-j_{1}(x)-1} + \mathcal{O}(y^{2}) \,,  
\label{eq:regge_exp}
\end{align}
where $j_{0} = -1 + {\mathcal{O}}(g^2)$ corresponds to the leading power in $y$, and $j_{1} = -2 +  {\mathcal{O}}(g^2)$ to the subleading power. 

Let us give explicit formulas for the functions appearing in eq. (\ref{eq:regge_exp}), up to two loops.
At the leading power, the residue function is given by
\begin{align}\label{eq:r0}
\begin{split}
r_{0}(x)=&1-g^4\Bigg\{\frac{2}{3}\frac{x(1-x)}{(1+x)^2}\big(\log^2(x)+3 \pi^2\big)\log^2(x) + \frac{1-x^2}{(1+x)^2}\bigg[ \frac{3}{10}\pi^{4} + \\
&2 \pi ^2 \text{Li}_2\left(x^2\right)+3 \text{Li}_4\left(x^2\right)+2 \text{Li}_2\left(x^2\right) \log ^2(x)-4 \text{Li}_3\left(x^2\right) \log (x)- \\
&4 \pi ^2 \text{Li}_2(x)+4 \zeta_{3} \log (x) \bigg] \Bigg\} \,,
\end{split}
\end{align}
and the leading Regge trajectory is given by
\begin{align}\label{eq:j0}
\begin{split}
j_{0}(x) = &-1 +2 g^2 \xi \log (x) - g^4 \Bigg\{ \frac{4}{3} \xi \log (x) \left(\log ^2(x)+\pi ^2\right)\\
& +\frac{2}{3} \xi^2  \Big[ 6 \text{Li}_3\left(x^2\right)-6 \text{Li}_2\left(x^2\right) \log (x)-2 \log ^3(x)-\pi ^2 \log (x)-6 \zeta_{3} \Big]
\Bigg\}  \,,
\end{split}
\end{align}
where we introduced $\xi = (1-x)/(1+x)$.
At the subleading order in $y$, the residue function is given by
\begin{align}\label{eq:r1}
r_{1}(x) = -2r_{0} + 4 + g^4 \frac{2}{3} \frac{(1-x)^2}{(1+x)^2}\Big( 12 \pi^2 +3 \pi^2 \log^{2}(x) +\log^{4}(x) \Big) \,,      
\end{align}
and the subleading Regge trajectory is
\begin{align}\label{eq:regge_sub_leading}
j_{1}(x) = -2 + g^4\frac{4}{3} \xi \Big( \pi^2 + \log^{2}(x) \Big) \log (x) \,.   
\end{align}
The following comments are in order:
\begin{itemize}
\item Eq. (\ref{eq:regge_exp}) means that up to subleading power in $y$, and up to a given loop order, all logarithms in $y$ are obtained from expanding the RHS of eq. (\ref{eq:regge_exp}) in the coupling $g^2$.
The fact that the subleading term can be written as a single power is non-trivial. It would be interesting to test this at higher loops. 
\item We have also looked at the sub-subleading power suppressed terms, which are of order $y^2$, but did not find an equally simple exponentiation pattern as for the leading and subleading terms. We leave this interesting topic to future studies.
\item As we discuss in subsection \ref{subsec:Wilsonloops}, remarkably, both the leading and subleading Regge trajectories are related to anomalous dimensions of certain Wilson loops.
\end{itemize}

\subsection{$SO(4)$ invariant Regge limit including $SO(6)$ angles}

The analysis above can be extended to involve $SO(6)$ angles, according to the AFHSTb setup.
While the loop functions remain unchanged thanks to dual conformal symmetry, this introduces a dependence on $\theta_{13}$ and $\theta_{24}$ into their normalizations, cf. section \ref{sec:angle_coulomb}.

In this case, we
find an analog of eq. (\ref{eq:regge_exp}), namely
\begin{align}
\begin{split}
\frac{1-y^{2}}{1+y^{2}-2y\cos(\theta_{24})} M(x,y,\theta_{13},\theta_{24}) \stackrel{{y \to 0} }{=} & r_{0}(x,\theta_{13},\theta_{24})y^{-j_{0}(x,\theta_{13})-1} \\
& + r_{1}(x,\theta_{13},\theta_{24})y^{-j_{1}(x,\theta_{13})-1} + \mathcal{O}(y^{2}) \,.  
\end{split}
\label{eq:regge_exp2}
\end{align}
The normalization on the LHS was chosen for convenience.
We find the following residue functions, up to two loops,
\begin{align}
\begin{split}
&r_{0}(x,\theta_{13},\theta_{24}) = 1 + g^{4}\xi\Bigg[ \frac{2}{3}\frac{x(x-\cos(\theta_{13}))}{(1-x^2)} \log ^2(x) \left(\log ^2(x)+3 \pi ^2\right) 
-3 \text{Li}_4\left(x^2\right) \\
&-2 \text{Li}_2\left(x^2\right) \left(\log ^2(x)+\pi ^2\right)+4   \Bigg( \text{Li}_3\left(x^2\right) 
- \zeta_3
\Bigg)
\log (x)
 +4 \pi ^2 \text{Li}_2(x)-\frac{3 \pi ^4}{10}
\Bigg] \,,
\end{split}
\end{align}
and 
\begin{align}
\begin{split}
&r_{1}(x,\theta_{13},\theta_{24}) = 2\cos(\theta_{24}) +g^{4} \xi 
\Bigg[ \frac{\cos (\theta_{24})}{15} \bigg(
60 \bigg(\text{Li}_2\left(x^2\right) \left(\log ^2(x)+\pi ^2\right)\\
& -2 \pi ^2 \text{Li}_2(x)\bigg)  
 -120 \text{Li}_3\left(x^2\right) \log (x)+10 \log (x) \left(\log ^3(x)+3 \pi ^2 \log (x)+12 \zeta_{3}\right)  \\ 
&+9 \pi ^4 +90 \text{Li}_4\left(x^2\right) \bigg) + \frac{4 \pi ^2 (1-x) (1+ \cos (\theta_{13}))}{(1+x)}
\Bigg] \,,
\end{split}
\end{align}
Remarkably, we find that the expressions for the Regge trajectories can be written in the same way as eqs. (\ref{eq:j0}) and (\ref{eq:regge_sub_leading}), as long as we replace $\xi$ by
\begin{align}\label{eq:xigeneral}
\xi = \frac{1+x^{2}-2x\cos(\theta_{13})}{1-x^{2}} \,.
\end{align}
We collect the expressions for $j_{0}$, $r_{0}$, $j_{1}$ and $r_{1}$ in the ancillary files \texttt{so(4)\_expansion\_ \\ ancillary.txt} and \texttt{so(4)\_expansion\_so(6)\_angle\_ancillary.txt} for zero and non-zero $SO(6)$ angles respectively.

\subsection{Regge trajectories and cusped Wilson loops with operator insertion}
\label{subsec:Wilsonloops}

As discussed in  \cite{Henn:2010bk,Correa:2012nk,Caron-Huot:2014gia}, the Regge limit of scattering amplitudes on the Coulomb branch of $\mathcal{N}=4$ sYM is closely connected to cusped Wilson loops. In QCD the same conenction holds for the massless limit only.

Indeed, the Regge trajectory $j_0$ given in eq. (\ref{eq:j0}), together with its dependence on $\theta_{13}$ according to eq. (\ref{eq:xigeneral}), is equivalent to the anomalous dimension of a Wilson loop with cusp angle $\phi$, and with an jump of $\theta_{13}$ in the coupling to the scalars between the two segments, cf. \cite{Drukker:2011za}.
The exact correspondence is \cite{Correa:2012nk}:
\begin{align}
   \Gamma_{\rm cusp}\left(x, \theta \, \right) =  - j_0(x, \theta) -1   \, .
\label{eq:regge_cusp}
\end{align}
In reference \cite{Bruser:2018jnc}, it was observed that the subleading Regge trajectory can be related to scaling dimension of a cusped Wilson line with a scalar insertion. In that paper, the scalar insertion corresponded to the same scalar that couples to the supersymmetric Wilson line. Here, we find a different expression for $j_1$, cf. eq. (\ref{eq:regge_sub_leading}), whose one-loop correction vanishes. 
We take this as a hint to consider instead the Wilson loop with an operator insertion that is orthogonal to the scalars the Wilson loop couples to, as this forbids one-loop corrections. 
At two loops, only a single Feynman diagram contributes.
The corresponding two-loop contribution to 
$\Gamma_{{\rm cusp}, \Phi}(x, \theta=0)$ can be found in Appendix A.2 of \cite{Correa:2012nk}. 
It is easy to promote this to general $\theta$.
Remarkably, we find that up to two loops,
\begin{align}\label{eq:j1_regge}
\Gamma_{{\rm cusp}, \Phi}(x,\theta) = -j_{1}(x,\theta) -1\,,     \end{align}
where $j_{1}$ is given by \eqref{eq:regge_sub_leading}, with the $\theta$ dependence entering via eq. (\ref{eq:xigeneral}).

We note that the LHS of eq. (\ref{eq:j1_regge}) is known in principle to any loop order: the anomalous dimension of such a cusped Wilson loop with $L$ scalar insertions has been computed in \cite{Correa:2012hh,Drukker:2012de,Gromov:2013qga} from integrability.
It would be very interesting to check formula \eqref{eq:regge_exp} at the three-loop level. 
This data could also help shed light on a possible general pattern for the 
second power suppressed terms, which is yet to be found.

\section{Conclusion and future directions}

In this paper we studied a four-point amplitude on the Coulomb branch of planar ${\cal N}=4$ theory. Our choice of VEV was motivated by the deformed Amplituhedron geometry and associated finite amplitude of ref. \cite{Arkani-Hamed:2023epq}. The VEV configuration chose here differs from the one discussed in \cite{Alday:2009zm}. In particular, it uses different $SO(6)$ directions in the VEV matrix of the scalar fields. We derived the corresponding Lagrangian of the theory identifying new mass and cubic interaction terms. 

In this setup we performed one-loop and two-loop calculations. Using generalized unitarity, we fixed the normalization coefficients of box and double box integrals (which are not determined by the Amplituhedron geometry). We evaluated the corresponding Feynman integrals analytically, using the four-dimensional equations method \cite{Caron-Huot:2014lda}.
As a main new result, we evaluated the Regge limit of the amplitude, including power suppressed terms. We found a pattern of exponentiation, analogous to that found in reference \cite{Bruser:2018jnc}, however with a different anomalous dimension governing the subleading power exponent. 
The exponent corresponds to a cusped supersymmetric Wilson loop that depends on both a geometric angle (given by the scattering kinematics) and an $SO(6)$ angle (corresponding to a parameter in our VEV setup).
For the reader's convenience, we provide results in computer-readable ancillary files 
containing the main formulas from sections \ref{sec:de} and \ref{sec:Regge}.

It would be very interesting to explore this connection further. The anomalous dimensions we identified in the Regge limit are known from integrability, and thus this turns into a prediction at higher loop orders. It would be interesting to test this prediction. Higher-loop data could also be invaluable in exploring a systematic pattern at the sub-subleading power in the Regge limit.

The main open conceptual question concerns the equivalence of the 
Coulomb branch theory and the deformed Amplituhedron geometry, which remains conjectural.  
At three loops, the deformed Amplituhedron geometry is more complicated compared to the massless case, but a preliminary analysis suggests that the canonical form can be written in terms of usual diagrams with deformed propagators, corresponding to the following two diagrams (and cyclic),
$$
    \includegraphics[width=7cm]{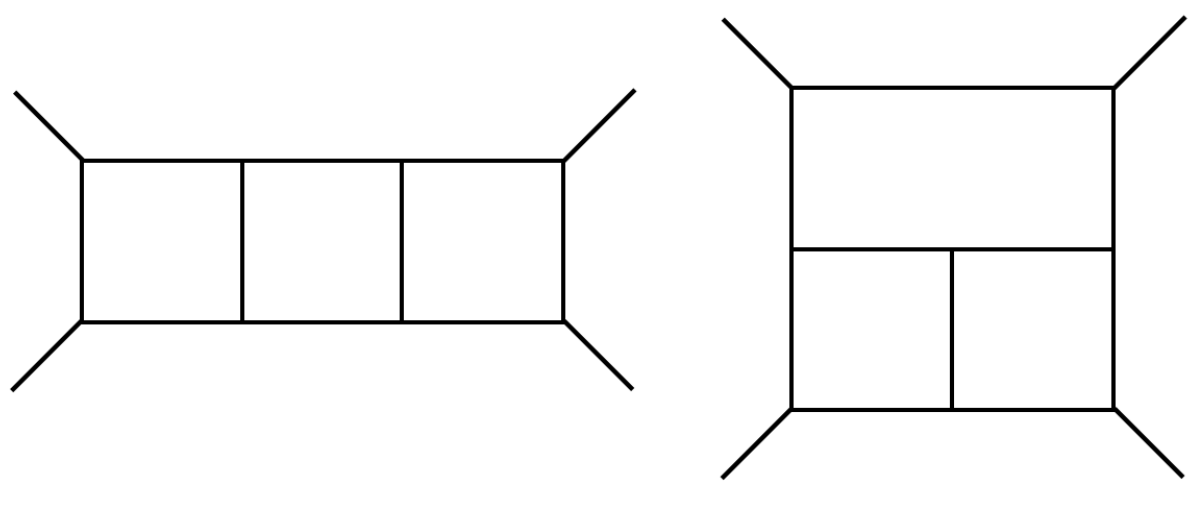}
$$
Although the singularities of the ladder integral are still logarithmic, the deformed propagators of the tennis court do not allow us to access any leading singularity (only 11 out of 12 parameters can be fixed on cuts). This suggests that integration would produce elliptic functions. This seems to be in conflict with the definition of the canonical {\it dlog} form for the deformed Amplituhedron geometry. The resolution of this puzzle is the most important immediate question.

Another important direction is the study of different Coulomb branch configurations and other (more general) deformations of the Amplituhedron. On the Coulomb branch side, we can include more independent $SO(6)$ vectors into the vacuum expectation values of the scalar fields or add intrinsic angles between these vectors. We can also extend the analysis to higher-point amplitudes where the space of Amplituhedron deformations is large and one needs to specify extra conditions on the deformed momentum-twistor variables to reduce the number of free parameters. Finally, it would be interesting to define an Amplituhedron-like geometry for one of the interesting kinematical limits, such as the Regge limit. Such a putative geometry should be dramatically simpler than the original space, reflecting simplifications in the kinematical expressions for amplitudes in these limits.

\bigskip
\section*{Acknowledgments}

It is a pleasure to thank Hofie Hannesdottir for discussions, and Simon Caron-Huot for corrrespondence. This work received funding from the European Union (ERC, UNIVERSE PLUS, 101118787). Views and opinions expressed are however those of the authors only and do not necessarily reflect those of the European Union or the European Research Council Executive Agency. Neither the European Union nor the granting authority can be held responsible for them. J.~T. is supported by the U.S. Department of Energy, grant No. SC0009999 and the funds of the University of California. W.~F. acknowledges the support of the ``Amplitudes'' INFN initiative.

\appendix

\section{Symbol alphabet and kinematics}\label{sec:appendix}

Here we provide useful formulas for switching between various parametrizations of the kinematics of the four-point amplitude considered in the main text.
The amplitudes depend on the variables $m_1, m_2, m_3, m_4, s, t$ via two dual conformal ratios only. 
The latter can be parametrized via two parameters $x,y$, as follows,
\begin{align}\label{varsuvxy}
 \frac{(-s+m_1^2 + m_3^2)^2}{4 m_1^2 m_3^2} = %
\frac{1}{4} \left(x + \frac{1}{x}\right)^2\,, \quad
 \frac{(-t+m_2^2 + m_4^2)^2}{4 m_2^2 m_4^2} = 
\frac{1}{4} \left(y + \frac{1}{y}\right)^2\,.
\end{align}
When inverting these relations, the following square root appears,
\begin{align}
\Delta(s,m_1^2,m_3^2) = \sqrt{ (s-m_1^2-m_3^2)^2-4 m_1^2 m_3^2} \,,
\end{align}
and similarly in the $t$-channel.
The inverse relation can be written as
\begin{align}\label{defxgeneral}
x &= \sqrt{ \frac{ m_1^2 +m_3^2 -s - \Delta(s,m_1^2,m_3^2) }{m_1^2 +m_3^2 -s + \Delta(s,m_1^2,m_3^2) }} \,, \\
y &= \sqrt{ \frac{ m_2^2 +m_4^2 -s - \Delta(t,m_2^2,m_4^2) }{m_2^2 +m_4^2 -s + \Delta(t,m_2^2,m_4^2) }} \,.\label{defygeneral}
\end{align}
We note that with this choice, in the Euclidean region $s<0, t<0, m_i^2>0$, we have the relations $0<x<1, 0<y<1$.
In the following, we provide useful formulas for switching between the symbol alphabet in $x,y$ variables and kinematic variables.

\subsection{Kinematics and alphabet for uniform mass case $m_{i} = m$}

For a uniform internal mass, we have $p_{i}^2 = 2 m^2$.
We then have independent variables
\begin{align}\label{varsnatural}
s,t,m^2 \,.
\end{align}
The relationship to the variables $x$ and $y$ is then,
\begin{align}
\frac{-s}{m^2} = \frac{(1-x)^2}{x} \,,\quad\quad \frac{-t}{m^2} = \frac{(1-y)^2}{y}\,.
\end{align}
Here we find a simpler expression for $x,y$ compared to the unequal mass case (again working in the Euclidean region $s<0,t<0, m^2>0$),
\begin{align}\label{inversionxyequalmasscase}
x=\frac{\sqrt{4 m^2 - s} - \sqrt{-s}}{\sqrt{4 m^2 - s} + \sqrt{-s}}\,, \qquad
y=\frac{\sqrt{4 m^2 - t} - \sqrt{-t}}{\sqrt{4 m^2 - t} + \sqrt{-t}}\,.
\end{align}
As before, these are chosen such that $0<x<1, 0<y<1$.
In terms of $x,y$, the alphabet consists of the following ten  letters,
\begin{align}\label{alphabetxy10lettersA}
\{ x, 1+x, 1-x, y, 1+y, 1-y, 1-x y, 1+x y ,x-y ,x+y \}\,.
\end{align}
We now provide relations to express those conveniently in terms of the variables (\ref{varsnatural}). We have 
\begin{align} \label{evenlettersstm}
\begin{split}
  \frac{-s}{m^2} =&  \frac{(1-x)^2}{x} \,,\quad  \frac{-s + 4 m^2}{m^2} =  \frac{(1+x)^2}{x}  \,,\quad \frac{4 m^2-s-t}{m^2} = \frac{(x+y)(1+x y)}{x y} \,,  \\
   \frac{-t}{m^2} =& \frac{(1-y)^2}{y} \,,\quad \frac{-t+4 m^2}{m^2}  =\frac{(1+y)^2}{y} \,,\quad \frac{ s-t}{m^2} =  \frac{(x-y)(1-x y)}{x y}  \,,
   \end{split}
\end{align}
and
\begin{align}\label{eq1alphabet}
\begin{split}
\frac{x-y}{1-x y} =&\frac{\sqrt{4 m^2 - s} \sqrt{-t}- \sqrt{4 m^2 - t}\sqrt{-s}}
{\sqrt{4 m^2 - s} \sqrt{-t}+ \sqrt{4 m^2 - t}\sqrt{-s}}\,,\\
\frac{x+y}{1+x y} =&\frac{\sqrt{4 m^2 - s}\sqrt{4 m^2 - t} - \sqrt{-t}\sqrt{-s}}
{\sqrt{4 m^2 - s}\sqrt{4 m^2 - t} + \sqrt{-t}\sqrt{-s}}\,.
\end{split}
\end{align}
Taken together, eqs. (\ref{inversionxyequalmasscase}),(\ref{evenlettersstm}), and (\ref{eq1alphabet}) allow us to rewrite the full alphabet (\ref{alphabetxy10lettersA}) in terms of $s,t, m^2$.
%
%

\subsection{Kinematics and alphabet for general mass case}

In the general mass case, $x$ and $y$ are given by eqs. (\ref{defxgeneral}) and (\ref{defygeneral}).
Here we provide compact expressions for the other alphabet letters.
For letters that depend on one variable only, we have
\begin{align}
\frac{-s+(m_1 - m_3)^2}{m_1 m_3} =& \frac{(1-x)^2}{x} \,, \qquad 
\frac{-s+(m_1 + m_3)^2}{m_1 m_3} = \frac{(1+x)^2}{x} \,,\label{lettergeneral4}\\
\frac{-t+(m_2 - m_4)^2}{m_2 m_4} =& \frac{(1-y)^2}{y} \,, \qquad 
\frac{-t+(m_2 + m_4)^2}{m_2 m_4} = \frac{(1+y)^2}{y} \label{lettergeneral6}\,.
\end{align}
Moreover, we have
\begin{align}
  \frac{-t +m_2^2+m_4^2}{m_2 m_4} -\frac{-s +m_1^2+m_3^2}{m_1 m_3}  =& \frac{(x-y)(1-x y)}{x y} \,,\\
  \frac{-t +m_2^2+m_4^2}{m_2  m_4} +\frac{-s +m_1^2+m_3^2}{m_1 m_3} =& \frac{(x+y)(1+x y)}{x y} \,.
\end{align}
Finally, we need two letters with square roots that generalize eq. (\ref{eq1alphabet}).
We find
\begin{align}
\frac{   (m_1^2+m_3^2-s) \Delta(t,m_2^2,m_4^2)+ (m_2^2+m_4^2-t) \Delta(s,m_1^2,m_3^2) } {2 m_1 m_2 m_3 m_4}  =&  \frac{(1-x y)(1+x y)}{x y} \,,\\
\frac{   (m_1^2+m_3^2-s) \Delta(t,m_2^2,m_4^2)-(m_2^2+m_4^2-t) \Delta(s,m_1^2,m_3^2) } {2 m_1 m_2 m_3 m_4}  =&  \frac{(x- y)(x+ y)}{x y} \,.
\end{align}

\bibliographystyle{JHEP}
\bibliography{refs}

\end{document}